\documentclass{aa} 
\usepackage{graphicx}
\usepackage[varg]{txfonts}
\usepackage{amsmath}

\def\name{\textsc{RefleX}}
\begin{document}

   \title{\name: X-ray absorption and reflection in active galactic nuclei
   for arbitrary geometries}

   \author{S. Paltani
          \inst{1}
          \and
          C. Ricci
          \inst{2,3,4}
          }

   \institute{Department of Astronomy, University of Geneva, Versoix, Switzerland\\
             \email{stephane.paltani@unige.ch}
         \and
             Instituto de Astrof\'{\i}sica, Facultad de F\'{i}sica, Pontificia Universidad Cat\'{o}lica de Chile, 306, Santiago 22, Chile
 \and Kavli Institute for Astronomy and Astrophysics, Peking University, Beijing 100871, China\\
\and Chinese Academy of Sciences South America Center for Astronomy and China-Chile Joint Center for Astronomy, Camino El Observatorio 1515, Las Condes, Santiago, Chile\\
\email{cricci@astro.puc.cl}
             }
   \titlerunning{\name: X-ray absorption and reflection in AGN for quasi-arbitrary geometries}

   \date{Received August 31, 2016; accepted May 19, 2017}

 
  \abstract{Reprocessed X-ray radiation carries important information about the structure and physical characteristics of the material surrounding the supermassive black hole (SMBH) in active galactic nuclei (AGN). We report here on a newly developed simulation platform, \name, which allows to reproduce absorption and reflection by quasi-arbitrary geometries. We show here the reliability of our approach by comparing the results of our simulations with existing spectral models such as \textsc{pexrav}, \textsc{MYTorus} and \textsc{BNTorus}. \name\ implements both Compton scattering on free electrons and Rayleigh scattering and Compton scattering on bound electrons. We show the effect of bound-electron corrections on a torus geometry simulated like in \textsc{MYTorus}.
  We release with this paper the \name\ executable, as well as \textsc{RXTorus}, a model that assumes absorption and reflection from a torus with a varying ratio of the minor to major axis of the torus. To allow major flexibility \textsc{RXTorus} is also distributed in three components: absorbed primary emission, scattered radiation and fluorescent lines. \textsc{RXTorus} is provided for different values of the abundance, and with (atomic configuration) or without (free-electron configuration) taking into account Rayleigh scattering and bound electrons. We apply the \textsc{RXTorus} model in both configurations on the XMM-Newton and NuSTAR spectrum of the Compton-thick AGN NGC\,424 and find that the models are able to reproduce very well the observations, but that the assumption on the bound or free state of the electrons has significant consequences on the fit parameters. 
  }


   \maketitle

\section{Introduction}

X-ray emission is one of the key characteristics of active galactic nuclei (AGN). Its existence is most probably the signature of the release of gravitational energy in a hot plasma in the vicinity of the supermassive black hole (SMBH) through Comptonization of soft photons, presumably from the accretion disk \citep[e.g.,][]{Haardt1993}. The study of direct X-ray emission is very interesting in itself, because of the complex physical processes at play. However, X-rays also provide a direct probe of the geometry of AGN through the study of their propagation through the surrounding matter which is irradiated by the central source. A fluorescence iron line at 6.4\,keV, corresponding to the K$\alpha$ transition, was discovered in early X-ray studies of AGN \citep[e.g.,][]{Mushotzky1978} and successively confirmed for a significant number of objects by EXOSAT \citep[e.g.,][]{Nandra1989,Pounds1989}. Studying the composite Ginga spectra of 12 AGN, \citet{Pounds1990} found evidence of a bump at $\sim 30$\,keV, the reflection hump. Both spectral features can be explained by the reprocessing of primary X-ray photons in the surrounding medium, and are therefore closely linked. While the Fe K$\alpha$ line can originate in Compton-thin material, the reflection hump implies the presence of material with significant optical depth to Compton scattering, i.e. close to being Compton-thick (CT; i.e. $N_{\rm\,H}\sim \sigma^{-1}_{\rm\,T}\simeq 1.5\times 10^{24}\rm\,cm^{-2}$).

Different structures around the supermassive black holes may contribute to the reprocessed X-ray radiation. An optically thick geometrically-thin accretion disk is supposed to channel matter onto the supermassive black hole. Moderately dense clouds \citep[$N_{\rm\,H}\sim 10^{23-24}\rm\,cm^{-2}$; e.g.][]{Netzer2015} in the broad-line region (BLR), with an emissivity-weighted radius spanning from a few to hundreds of light days \citep[depending on the AGN luminosity, e.g.,][]{Kaspi2005} are thought to be responsible for the emission of broad lines in the infrared to ultraviolet. More distant and rarefied matter (the narrow-line region or NLR), with densities of $n_{\rm\,e}\sim 10^{3}-10^{7}\rm\,cm^{-3}$ \citep[e.g.,][]{Baskin2005} is responsible for the emission of narrow lines in the optical and infrared. At larger distances, the accretion disk is thought to become thick and form a torus, with the BLR marking possibly the transition between the two; absorbed AGN ($N_{\rm\,H}\geq 10^{22}\rm\,cm^{-2}$) are generally thought to be observed through the torus, although a contribution might come also from the BLR or from material located on galactic scales \citep[e.g.,][]{Goulding2012}. Finally, evidence for a fully ionized plasma has been found in extremely obscured objects in the form of a scattered component accounting for a few percent of the primary emission \citep{Turner1997}.

The physics of the interaction of X-ray photons with the surrounding material includes several processes. The most important processes are photoionization and elastic or inelastic scattering.  Photoionization is furthermore followed by the emission of either an electron (the Auger effect), a fluorescence photon or both (the Coster-Kronig process). The physics of these processes is well understood, and several reflection models are currently available in the standard \textsc{xspec} spectral analysis package \citep{Arnaud1996}. For instance, \textsc{pexrav} and \textsc{pexmon} assume the existence of optically thick, cold material (although it is assumed that H and He are fully ionized), distributed in a slab and covering a given fraction of the X-ray source. The \textsc{pexrav} model \citep{Magdziarz1995} ignores fluorescence, while \textsc{pexmon} \citep{Nandra2007}, adds a number of important spectral features, such as the fluorescence lines Fe\,K$\alpha$, Fe\,K$\beta$, and Ni\,K$\alpha$, following the Monte Carlo calculations of \citet{George1991} and the Compton shoulder (in the form of a Gaussian line of width $\sigma=35$\,eV centered at 6.315\,keV). While these models are still very popular, they have strong limitations. One of these limitations is that the reflective medium is completely optically thick, so transmission through the medium is totally suppressed. The geometry is limited to a planar geometry, so they are not applicable to realistic geometries beyond the accretion disk. They also assume that the source is very close to the reflecting medium, so that the incident angle of the primary radiation is isotropically averaged.

To overcome these limitations, new models have been developed. {\sc MYTorus}\footnote{Available at http://www.mytorus.com/} \citep{Murphy2009} implements a toroidal geometry with arbitrary column density. \citet{Ikeda2009} and \citet{Brightman2011} have developed models that consider a spherical toroidal structure with varying covering factors and constant $N_\mathrm{H}$ in all directions across the torus. All these torus models are based on Monte Carlo simulations, where primary X-ray photons are propagated through the simulated torus structure, undergoing physical processes along the way, until they are either destroyed through photoionization or they escape the system.

While these models represent a major advance over the traditional slab models, they still do not allow very flexible geometries, or even the self-consistent combination of simple geometries. For instance, in a model of a disk and a torus represented with \textsc{pexrav+mytorus}, photons reflected by the disk are unaware of the presence of the torus. We introduce here a new tool, \name, which is able to simulate theoretical X-ray spectra of AGN using quasi-arbitrary geometries. We present in this first paper the implementation of our simulation tool. We demonstrate the validity of our approach by comparing with standard and more recent transmission/reflection models. With this paper, we release publicly the \name\ executable\footnote{\name\ executable, user manual and example models are available at http://www.astro.unige.ch/reflex}. We provide in addition a series of \textsc{xspec} models, \textsc{RXTorus} (see Sect.\ref{sec:rxtorus}), generated with \name, which are extensions of the \textsc{MYTorus} with varying covering factors and metallicities. The goal is to compute and distribute regularly new X-ray spectral models to the community, adapting the geometry of the reprocessing material to different physical scenarios. In spite of the loss of the Soft X-ray Spectrometer \citep[SXS;][]{Mitsuda2014} on board Hitomi \citep{Takahashi2014}, the spectacular results of the SXS \citep{Hitomi2016} have demonstrated the need for similar instruments in the future. Therefore, a particular emphasis of our models is to be applicable to future-generation high-resolution X-ray instruments, such as the X-ray Integral Field Unit \citep[X-IFU;][]{Barret2016} on board the future large X-ray observatory of the European Space Agency Athena \citep{Nandra2013}.

\section{Ray-tracing simulations of X-ray transmission and reflection}

We developed a ray-tracing code, \name, with the aim of being able to simulate realistically the physical processes of propagation of X-rays, which we define as photons with energies from about 0.1\,keV to about 1\,MeV) through matter around the central engine of AGN. The \name\ code uses Monte Carlo simulations to track individual photons. The simulated photons interact with the matter through all the physical processes selected by the user (with sensible defaults), until the photon dies, is downscattered to an energy below a chosen termination energy or exits the system. 

The ray-tracing simulation consists of four separate engines that are closely connected: photon creation, geometrical description, photon propagation, and physical processes. The details of these engines are presented below.

The code records the full list of events that are experienced by each photon, although the level of detail is configurable to allow the user to manage the quantity of information produced. This allows us, for instance, to select photons that crossed a particular object or experienced a particular physical process. The time of flight of the photon is also recorded, allowing the user to perform timing studies.

The simulation is geometrically bounded by the \textsc{World}, which is a sphere with a user-defined radius centered on the origin. All processes occurring outside of \textsc{World} are ignored and ray tracing stops when a photon reaches \textsc{World}.

\subsection{X-ray generator}

The first step of the \name\ Monte Carlo simulation is the creation of photons. A photon is determined by a time, position, direction, and energy. All photons are created at time $t=0$. The source can be placed anywhere within the boundary of the simulation (\textsc{World}). The position and directions are drawn at random from some predetermined distributions and geometries selected by the user. The source can currently be a point, sphere, or disk. For each geometry, the photon direction can be either fully isotropic, or restricted to a sector. 

The photon energy is drawn from several possible distributions. These distributions are currently: monoenergetic, Gaussian, blackbody, power law, or power law with a high-energy exponential cutoff. New distributions can be easily implemented into \name.

\subsection{Objects}

The main novelty of \name\ is the possibility to define complex geometries for the surrounding material by assembling independent building blocks. An arbitrary number of objects can be set up by the user. Each object is characterized by a position, geometrical shape, composition, and hydrogen density. Four different geometries are currently included: sphere, disk or cylinder, annulus, shell, and torus. Annuli and shells can be used to simulate disks and spheres with density profiles. The objects can be placed anywhere within \textsc{World}, with a few exceptions: the axis of symmetry of disks, annuli and, tori must be parallel to the $z$-axis. These limitations could be removed in principle, at the price of more complex analytic treatment of the interactions (see Sect.~\ref{sec:prop}), if deemed useful. The objects can also be named, so that they can be identified in the tracking of the events.

Once a geometrical object has been defined, its composition needs to be set. A composition is defined by the number density ratios of all elements from $Z=2$ (helium) to $Z=30$ (zinc) compared to hydrogen. Each object can have a different composition and hydrogen number density. Several compositions have been predefined similarly to those available in \textsc{xspec}. These are \textsc{angr} \citep{Anders1989}, \textsc{aspl} \citep{Asplund2009}, \textsc{feld} \citep[][except for elements not listed that are given \textsc{grsa} abundances]{Feldman1992}, \textsc{aneb} \citep{Anders1982}, \textsc{grsa} \citep{Grevesse1998}, \textsc{wilm} \citep[][except for elements not listed that are given zero abundance]{Wilms2000} and \textsc{lodd} \citep{Lodders2003}. The metallicity (abundance of elements with $Z>2$) can be modified and the number density ratio of each element can be tuned individually to simulate arbitrary chemical compositions.

\subsection{Propagator}
\label{sec:prop}
Once a photon is created by the X-ray generator, the propagator tracks the photon, propagates it through the objects and determines the occurrence of physical interaction between the photon and the material contained in the objects. For this reason, the propagator always knows in which object the photon resides, and is aware of all interaction cross-sections between the photon and the object in which it resides.

The propagator analytically finds all intersections between the photon trajectory and the objects in order to determine which object is next. Each physical process is then tested sequentially (see Sect.~\ref{sec:phys}) to determine if any of these processes takes place inside the object, or if the photon exits the object. In all cases, the photon is first propagated to the location where the interaction takes place, or at the surface of the object if it exits. When an interaction takes place, the trajectory, intersections and cross-sections are recomputed, and the propagator is called again.

By geometrical construction, the propagator always finds one intersection in the future with \textsc{World}. Once the photon hits \textsc{World}, the propagation is terminated and the interaction parameters (location, direction, energy, and time) are recorded. The propagation also terminates if the photon is destroyed or if the photon energy drops below a minimum energy set by the user.

\subsection{Main physical processes}
\label{sec:phys}
The \name\ code implements several physical processes that take place when an X-ray photon crosses matter. We do not attempt to simulate all physical processes, but restrict ourselves to the few that are the most significant, similar to what has been the case in other models such as \textsc{pexrav} \citep{Magdziarz1995}, \textsc{pexmon} \citep{Nandra2007} and \textsc{MYTorus} \citep{Murphy2009}. The modular nature of the code allows us to add new physical processes if needed. In order to provide insight into the physical mechanisms, each process can be turned on and off and the processes that have taken place are recorded or each simulated photon. 
\subsubsection{Compton and Rayleigh scattering}
\label{sec:scattering}
Compton scattering is the inelastic (incoherent) interaction between a photon and a free electron. In the low-energy limit, the scattering becomes quasi-elastic, and is approximated by the Thomson scattering. The Compton scattering cross-section is given by the Klein-Nishina relativistic differential cross-section. The \name\ code implements the full Klein-Nishina cross-section for free electrons, although the user can disable both the Klein-Nishina correction and inelastic scattering, for educational purposes. Several models, such as \textsc{pexrav} \citep{Magdziarz1995}, assume that all H and He atoms are fully ionized, and scattering occurs only on these free electrons, neglecting the electrons bound on metallic atoms. The same physical configuration can be selected by the user in \name; in the following, we refer to this configuration as the free-electron configuration. 

However, electrons bound to atoms have a different behavior with respect to scattering of photons. The \name\ code can therefore adopt a different physical configuration, that we refer to as the atomic configuration, which takes into account the bound state of electrons. We include in \name\ binding correction terms for Compton scattering, called incoherent scattering functions, computed by \cite{Hubbell1975} as an energy- and $Z$-dependent multiplicative term on the Klein-Nishina cross-section. These corrections are also available in the EPDL97 library\footnote{https://www-nds.iaea.org/epdl97/}, and strongly decrease the cross-section for Compton scattering at low energy, which is essentially replaced by coherent (Rayleigh) scattering. 

Rayleigh scattering is a coherent scattering on atoms due to their polarizability. Contrarily to Compton scattering, the atom is neither ionized nor excited. Rayleigh scattering cross-sections are significantly larger than the Compton scattering cross-sections at low energy, up to several tens of keV for heavy atoms. The angular distribution is identical to that of Compton scattering in the low-energy limit. In the atomic configuration, \name\ implements Rayleigh scattering, unless disabled by the user. The Rayleigh cross-sections have been taken from \cite{Hubbell1975}.

Coherent and incoherent scattering on molecular hydrogen, H$_2$, is also included in \name\ in the atomic configuration using the cross-sections of \cite{Hubbell1975}. The fraction of hydrogen in molecular form can be set in \name.

Several recent models already include these corrections for scattering on bound electrons \citep[][see Sect.~\ref{sec:geant4}]{Liu2014,Furui2016}. Applying these corrections has been shown to make a difference of up to 25\% in the reflected spectrum at 1\,keV, the effect becoming negligible around and above 10\,keV \citep{Liu2014}.  

\subsubsection{Photoelectric absorption}
Photoelectric absorption by bound electrons is another important process affecting transmission and reprocessing of X-rays. We use here the cross-sections determined from the analytic fits for each shell of the 30 elements provided by \citet{Verner1995} and \citet{Verner1996}\footnote{These cross-sections are available in \textsc{xspec}, for instance in the \textsc{phabs} model, with the command \textsc{xsect vern}}; the cross-sections are computed using the Hartree-Dirac-Slater formalism and are thoroughly checked with existing laboratory measurements. In the current implementation, only neutral atoms are considered, although the analytic representations cover all ionization stages. 

The use of analytic functions allows us to avoid performing any interpolation in functions that have sharp edges. The cross-sections are calculated on the fly when the photon is created or when its energy changes, for instance after Compton scattering, at the precise photon energy. The cross-sections are calculated separately for each shell, which allows us to keep track of the shell responsible for the photoelectric absorption.

\subsubsection{Recombination}
When photoionization occurs, an electron is kicked out of the atom. The vacancy is filled by another electron from an outer electronic shell. Usually, the excess energy is transmitted to an outer-shell electron, which is then ejected. No further photon is emitted, and the simulation simply terminates. This process is known as the Auger effect.

In other cases, a fluorescence photon is emitted at an energy equal to the difference between the outer-shell energy level of the electron having filled the gap and the energy level of the vacancy. It is emitted in a random direction at the location of the interaction. The probabilities of such transitions, the fluorescence yields, which depend strongly on the atomic number of the element, have been taken from LBNL X-ray Data Booklet \citep{Thompson2009}; in particular, the Fe\,K$\alpha$/Fe\,K$\beta$ ratio is 8.147. The energy levels of a number of X-ray fluorescence lines have been measured in the laboratory by \citet{Bearden1967} and are therefore more accurate than the LBNL ones; we use these measured energies when available. Only fluorescence from K and L shells has been implemented. The L-shell fluorescence lines are considerably weaker and have energies below 1.2\,keV, but some lines are astrophysically relevant, as iron L-shell fluorescence has already been detected in an AGN \citep{Fabian2009}. The full list of K-shell fluorescence lines with the adopted fluorescence yields and energy levels is presented in Appendix~\ref{sec:fluotable}.

\subsection{Polarization}
In the context of reflection of X-rays on cold matter, photon polarization essentially impacts the direction in which incoming photons are scattered in Compton scattering. The formalism is presented in \citet{Matt1996}. While there are not many observational constraints on polarization in the X-rays, this could change in the future with the advent of X-ray polarimeters \citep[e.g.,][]{Soffitta2013}. Polarization-dependent Compton scattering in the Klein-Nishina regime is implemented as an option in \name. The study of polarized reflection with \name\ will be deferred to another paper.

\label{sec:pexrav}
\begin{figure}[tbp]
\includegraphics[width=8.8cm]{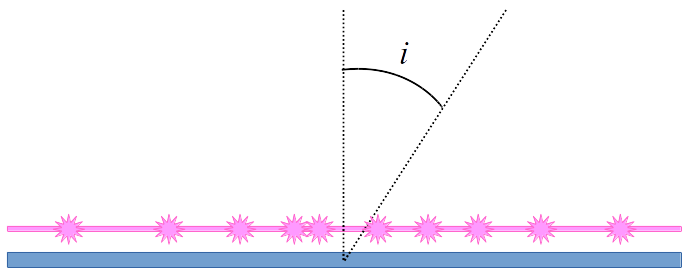}\\
\caption{\label{ill:pexrav}Geometry of the \textsc{pexrav}/\textsc{pexmon} model. The reflector is the blue plane. The source of X-ray photons is the magenta plane, parallel to the reflector. The inclination angle $i$ is defined as the angle between the normal to the  planes and the observer. The stars illustrate that primary photons are emitted isotropically anywhere in the source plane. Half of the primary photons leave the system without hitting the reflector.}
\end{figure}

\subsection{Normalization of spectra}
Since \name\ generates a fixed, user-selected number of photons $N$, the production of spectra in physical units of flux, either in counts s$^{-1}$ cm$^{-2}$ keV$^{-1}$ or in keV s$^{-1}$ cm$^{-2}$ keV$^{-1}$, requires a normalization factor. This is calculated by setting the luminosity of the unabsorbed source in a given energy range and its redshift; the redshift is in addition applied to all photons escaping to \textsc{World}.

We point out however that the parameter is the unabsorbed luminosity of the primary source of X-ray photons. In addition to the obvious effect of the absorber, two effects impact the observed luminosity. First, some part of the primary emission is reflected, thereby enhancing the observed luminosity; second, in the case of non-isotropic geometry, the flux and luminosity depend on the direction of the observer. The first effect should not be very important, since the reflected flux is generally a small fraction of the unabsorbed primary emission. The second effect may however be important in situations where the primary source is strongly anisotropic.

\subsection{Domain of validity}
The physical processes implemented in \name\ are exact, i.e. no simplifying assumption is made on the calculation of any of the physical processes described in this section. However, the accuracy of \name\ simulations is limited by a number of factors. Implemented physical processes depend on parameters which are usually only known with some limited precision and accuracy. As an example, the photoelectric cross-sections used in \name, while commonly used in astronomy, are not unique; in addition to the cross-sections used in \name\ (\textsc{vern}), \textsc{xspec} can also use the cross-sections from \citet{Balucinska1992} with either the new H and He cross-sections from \citet{Yan1998} (\textsc{bcmc}) or the original cross-sections (\textsc{obcm}).

Similarly, while the Klein-Nishina cross-section is computed using the exact formula, corrections for bound electrons and Rayleigh scattering cross-sections are computed using approximations, Compton and Rayleigh cross-sections in \name\ are interpolated using the tables from \citet{Hubbell1975}, which are valid from 100\,eV to 100\,MeV.

Many of the fluorescence lines (and all the most prominent ones) have been measured in the laboratory by \citet{Bearden1967}; however, the energy levels of many minor lines have been calculated. More importantly, all probability transitions are also calculated.

The main limitations of \name\ come however from the physical processes that are not implemented, in particular the following:
\begin{itemize}
\item Pair production contributes to opacity at energies above 1 MeV, and even starts to dominate around 10\,MeV. Likewise, photon-photon opacity is not included.
\item Matter velocities, thermal broadening and turbulence are not included, although we plan to include these processes at a later stage; this may induce unrealistically sharp features in the \name\ spectra. This is mitigated by the limited spectral resolution of X-ray telescopes.
\item Coster-Kronig process is a special case of an Auger effect where the vacancy is filled by an electron from the same shell. In this case, a secondary recombination may occur, which could be in the form of a fluorescence line. Since Coster-Kronig process cannot affect the K shell, only L-shell fluorescence and above can occur. Because L-shell fluorescence is already weak and because Coster-Kronig process is a second-order effect, it has not been implemented in \name. Likewise, M-shell (and above) fluorescence is neglected.
\item Raman scattering, which is an inelastic scattering of a photon with an atom or a molecule, is not included, since it has much lower probability than Rayleigh scattering.
\end{itemize}
\section{Comparison with existing models}
\label{sec:models}
Before applying our code to simulate the reprocessed emission of AGN, it needs to be validated. We therefore computed detailed simulations that mimic models available in \textsc{xspec} or in the literature.

\begin{figure}[h!]
\includegraphics[height=8.8cm,angle=-90]{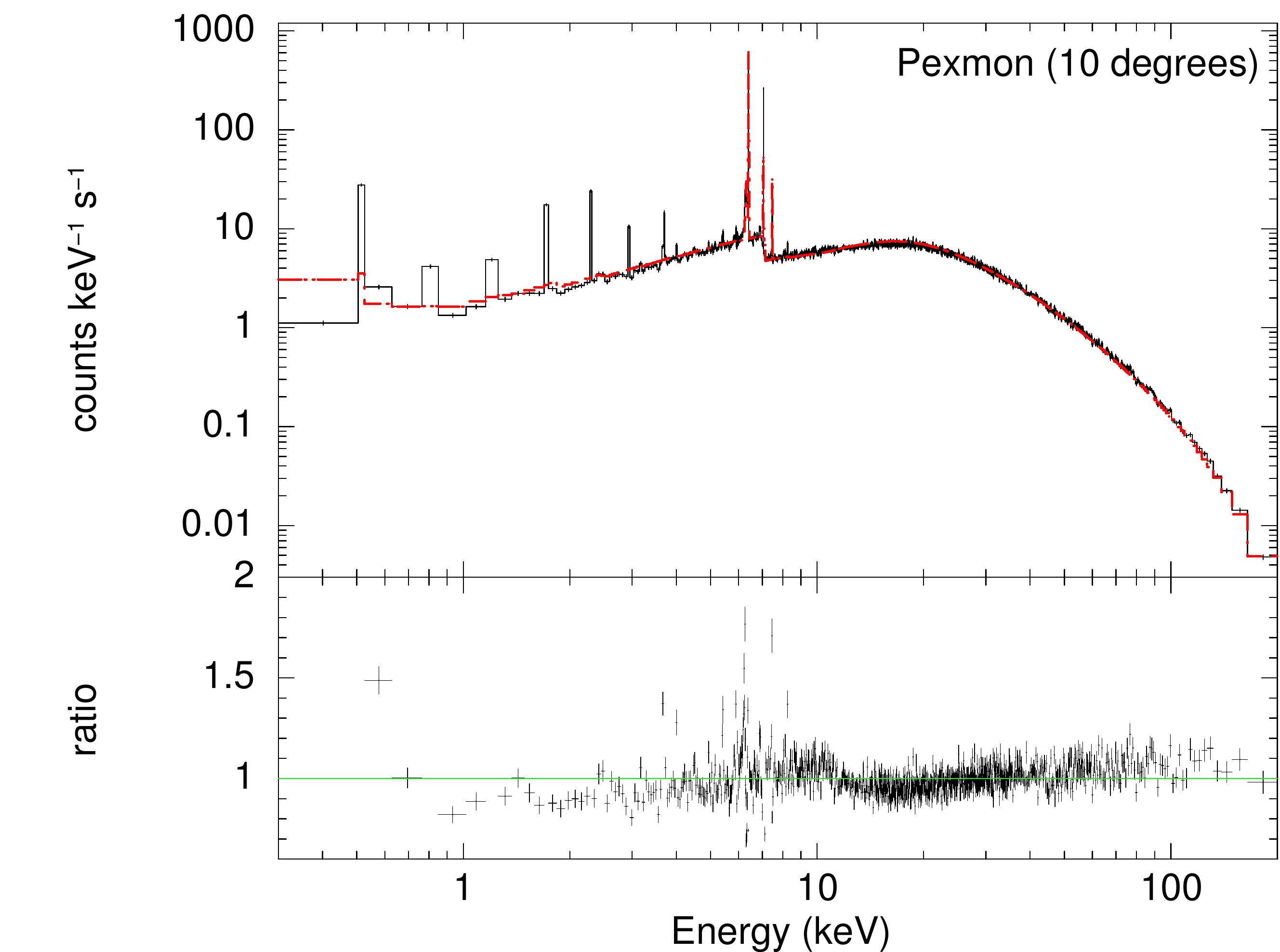}\\
\includegraphics[height=8.8cm,angle=-90]{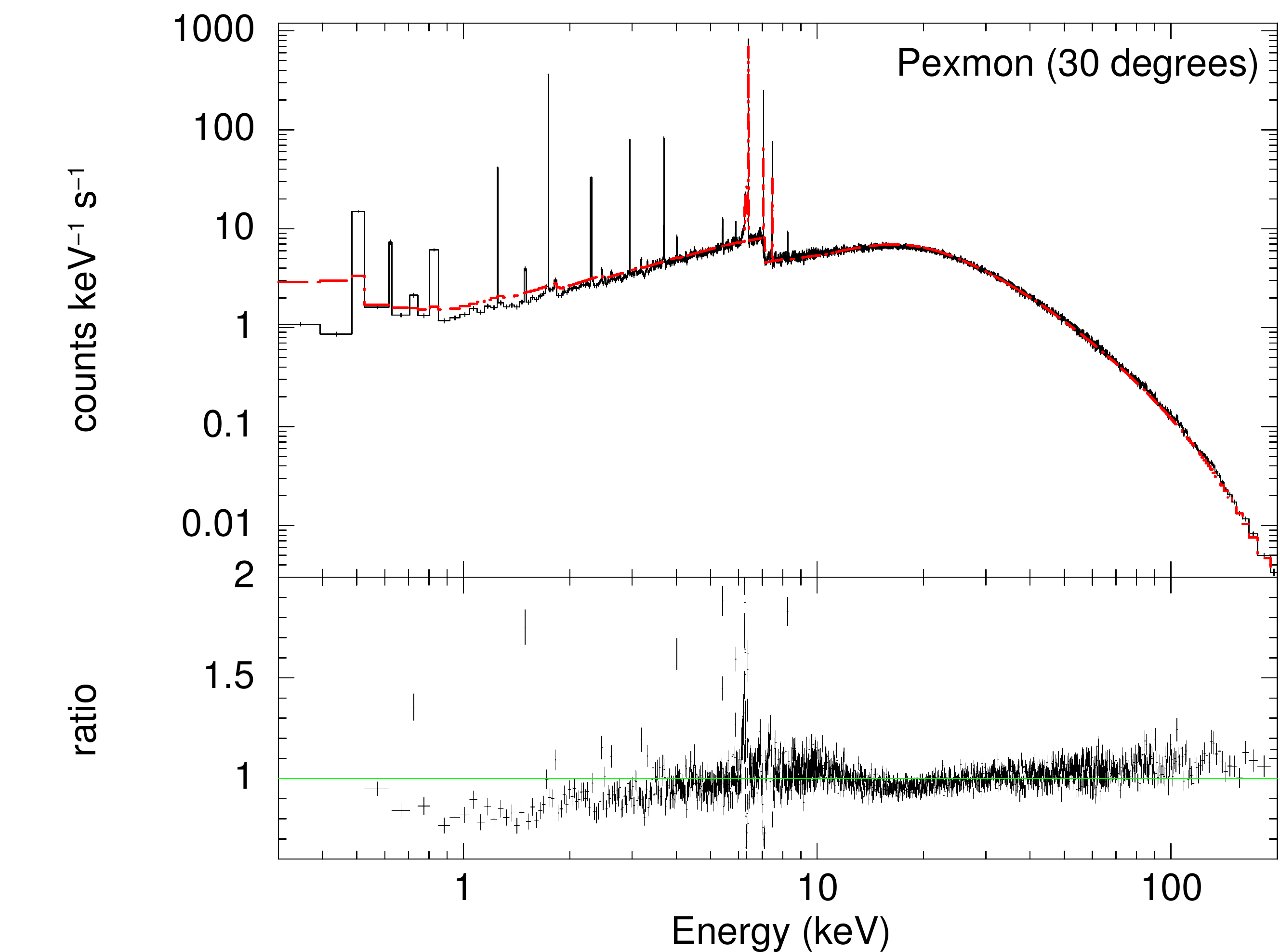}\\
\includegraphics[height=8.8cm,angle=-90]{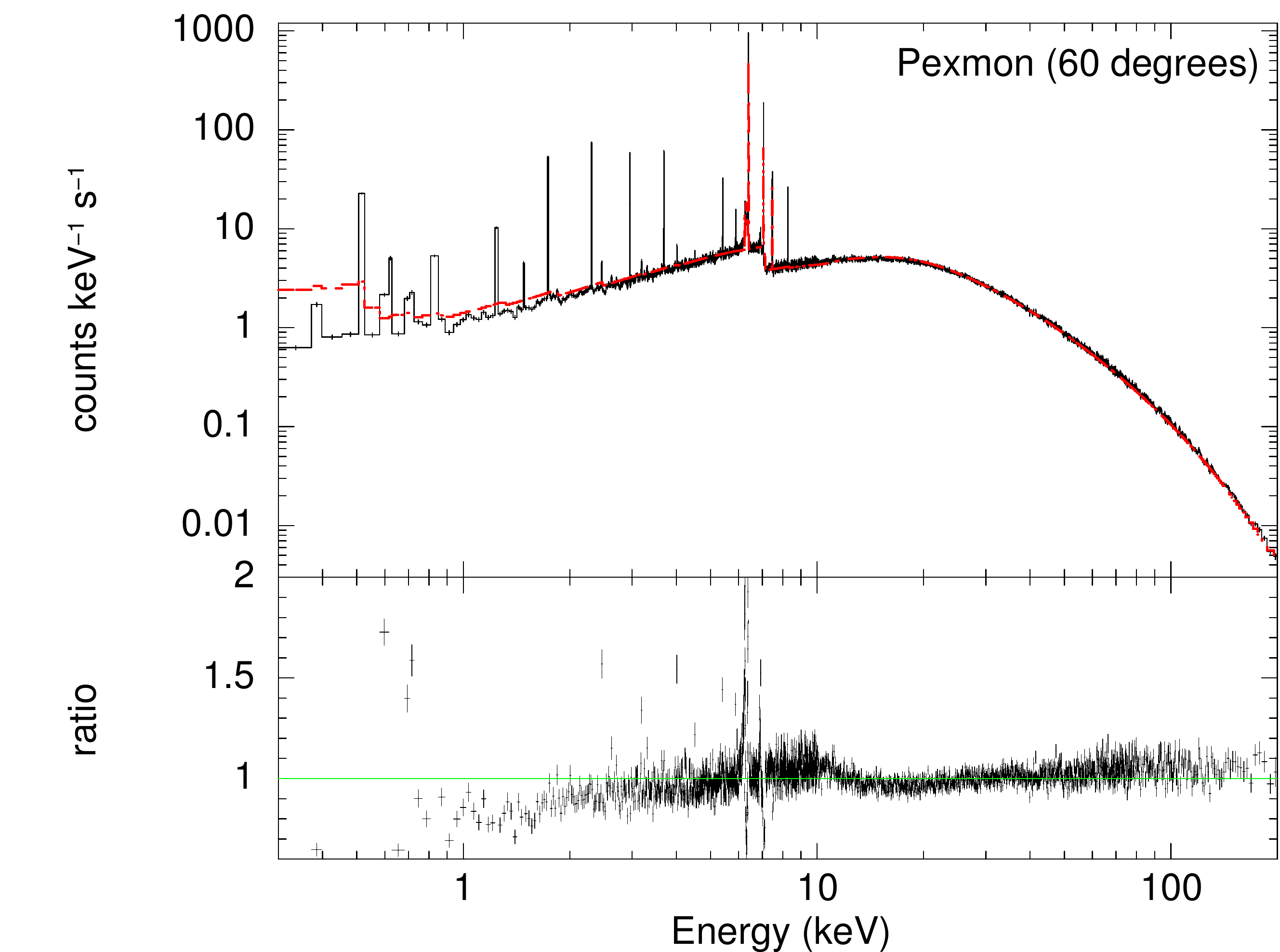}
\caption{\label{fig:pexmondeg}Comparison between \textsc{pexmon} and our \name\ simulations assuming the geometry of Fig.\,\ref{ill:pexrav} for three viewing angles. Top: $i=10$\,deg; middle: $i=30$\,deg; and bottom: $i=60$\,deg. In each panel, the top subplot shows the \textsc{pexmon} model (solid line) and the \name\ simulation (crosses); the bottom subplot shows the ratio between our simulations and the \textsc{pexmon} model. In all cases, the primary continuum is defined as a cutoff power law starting at 1\,keV with an index $\Gamma=1.5$ and a cutoff energy at 100\,keV. Spectra are averaged over 2 deg in inclination. $10^{10}$ photons were generated.}
\end{figure}
\begin{figure}[htbp]
\includegraphics[width=8.8cm]{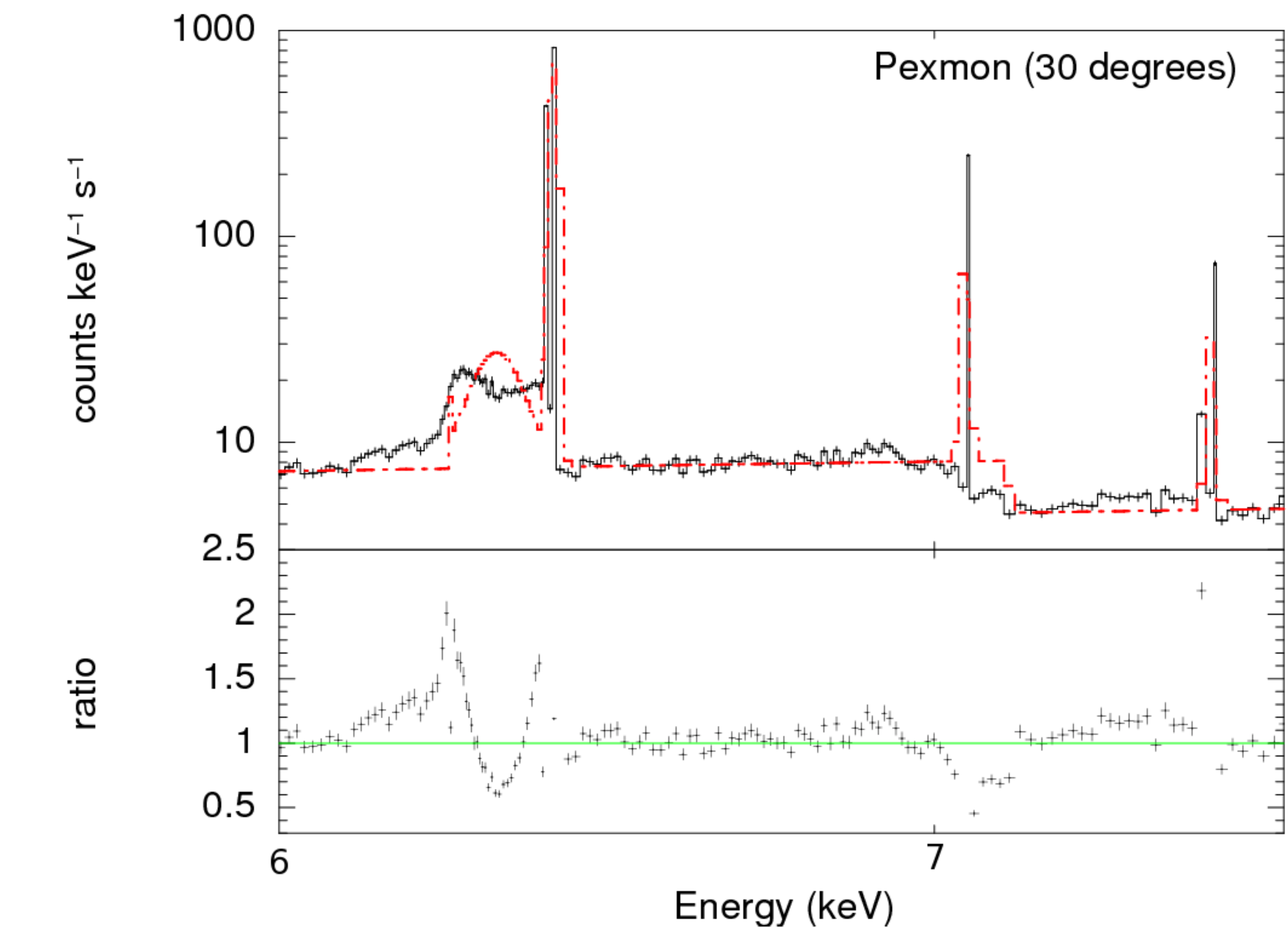}
\caption{\label{fig:pexmonzoom}Comparison between \textsc{pexmon} and our \name\ simulations assuming the geometry of Fig.\,\ref{ill:pexrav} for $i=30$\,deg, limited to the 6--7.6\,keV region. The top subplot shows the \textsc{pexmon} model (solid line) and the \name\ simulation (crosses); the bottom subplot shows the ratio between our simulations and the \textsc{pexmon} model. The primary continuum is the same as in Fig.~\ref{fig:pexmondeg}.}
\end{figure}

\subsection{pexrav and pexmon}

\textsc{pexrav} \citep{Magdziarz1995} is the standard reflection model in \textsc{xspec}. This model, which has been computed using analytic approximations of Green's functions derived from Monte Carlo simulations, assumes that the reflector is an optically thick medium in an infinite plane geometry and that the source is located immediately on top of the reflector; the extension of the source is irrelevant as long as it is significantly smaller than extension of the reflector. The distribution of incident angles follows $\mu=\cos \vartheta$, where $\vartheta$ is 0 for an incident photon perpendicular to the plane. Fig.\ \ref{ill:pexrav} illustrates the geometry adopted in \textsc{pexrav}. The primary X-ray emission follows a power law with an exponential cut off, although the \textsc{reflect} convolution model can be used to simulate the same physical processes using an arbitrary primary X-ray emission spectrum. \textsc{pexrav} implements photoelectric absorption and Compton scattering, including Klein-Nishina cross-section. \textsc{pexrav} assumes that all electrons from hydrogen and helium atoms are free, and disregards the other electrons. \textsc{pexmon} \citep{Nandra2007} is based on the \textsc{pexrav} model and implements the same geometry, but adds the Fe\,K$\alpha$ fluorescence line based on the Monte Carlo simulations by \citet{George1991}. Fe\,K$\beta$ and Ni\,K$\alpha$ fluorescence lines have also been added and their fluxes are fixed at a fraction of the Fe\,K$\alpha$ flux. In addition, a Compton shoulder caused by Fe\,K$\alpha$ fluorescence line that is scattered and shifted to lower energies has been added in the form of a Gaussian, following the prescription of \citet{Matt2002} regarding its intensity (but not its shape). These models depend on composition, metallicity, inclination angle $i$ of the disk with respect to the observer and on a normalization factor $R$, with $R=1$ indicating that the reflector covers half of the solid angle as seen from the source, i.e. it is quasi-infinite.

Figure~\ref{fig:pexmondeg} shows the comparison between \textsc{pexmon} and our simulations for different inclination angles ($i=10$, 30 and 60\,deg) and $R=1$. The primary continuum is defined as a cutoff power law  with an index $\Gamma=1.5$ and a cutoff energy at 100\,keV. The termination energy was set to 300\,eV. We selected the element abundances from \citet{Lodders2003} in both models and used the same photoionization cross-sections. We find that the shapes of the two spectra match generally well, although there are some significant differences. We observe that the reflected spectrum is slightly harder, in particular below $\sim$3--4 keV, which introduces a $\sim$20\% difference in flux at 1 keV. Another difference, which was expected, is the presence of several other fluorescence emission lines in our simulations, which are not implemented in \textsc{pexmon}. These lines have been detected in the Compton-thick AGN in \object{M51} \citep{Xu2015}. Finally, we point out the presence of an excess in the \textsc{pexmon} spectrum below $\sim0.5$\,keV. While this energy corresponds to that of the 1s edge of oxygen, such a feature is not really expected. We believe it is due to an imperfect modeling in \textsc{pexrav/pexmon}. With the exception of these differences, we find that the overall normalizations of the two spectra are in excellent agreement. The observed discrepancies are probably related to the approximations adopted in \textsc{pexrav}. 

\begin{figure}[tbp]
\includegraphics[width=8.8cm]{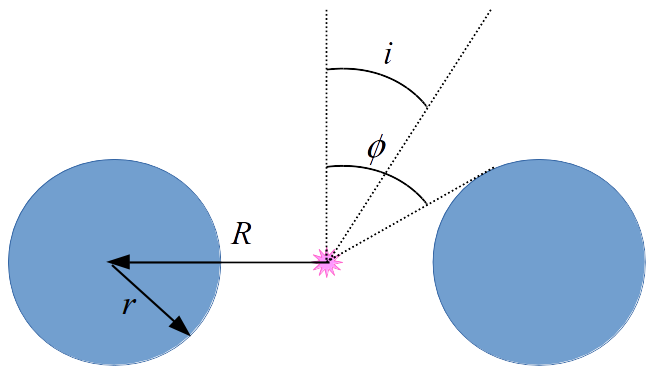}\\
\caption{\label{ill:mytorus}Geometry of the \textsc{MYTorus} model. The reflector is a torus, whose section is shown in blue. The source of X-ray photons is the magenta star in the center, which is isotropic. The inclination angle $i$ is defined as the angle between the normal to the  plane of the torus and the observer. In the \textsc{MYTorus} model, the ratio between the major and minor radii of the torus, $R/r$, is 2, leading to a half opening angle $\phi=60$\,deg.}
\end{figure}

Figure~\ref{fig:pexmonzoom} shows a zoom on the 6--7.6\,keV region for $i=30$\,deg. The Gaussian shape of the Compton shoulder implemented in \textsc{pexmon} does not match the results of our simulation at all. The latter is however perfectly similar to the Compton shoulder simulations of \citet{Matt2002}, which uses a similar Monte Carlo approach. In addition, the simulation includes separate fluorescence lines for the L$_2$-K$_1$ and L$_3$-K$_1$ transitions Fe\,K$\alpha_2$ and Fe\,K$\alpha_1$, respectively, whose energies differ by 13\,eV (6390.84 vs 6403.84\,eV respectively). These spectral features could be easily resolved by X-ray telescopes equipped with micro-calorimeters, such as Athena. The Fe\,K$\beta$ and Ni\,K$\alpha$ lines also appear different. This is due to the assumptions on the energy of these lines (7050\,eV and 7470\,eV, respectively) and on the width of these lines (35\,eV) in \textsc{pexmon}, while \name\ uses the values of \citet{Bearden1967}, i.e., Fe\,K$\beta$ at 7057.98\,eV and two separate Ni\,K$\alpha_2$ and Ni\,K$\alpha_1$ at 7460.89\,eV and 7478.15\,eV, respectively, and a 0\,eV line width.

\subsection{MYTorus}

\begin{figure*}[tbp]
\includegraphics[height=8.8cm,angle=-90]{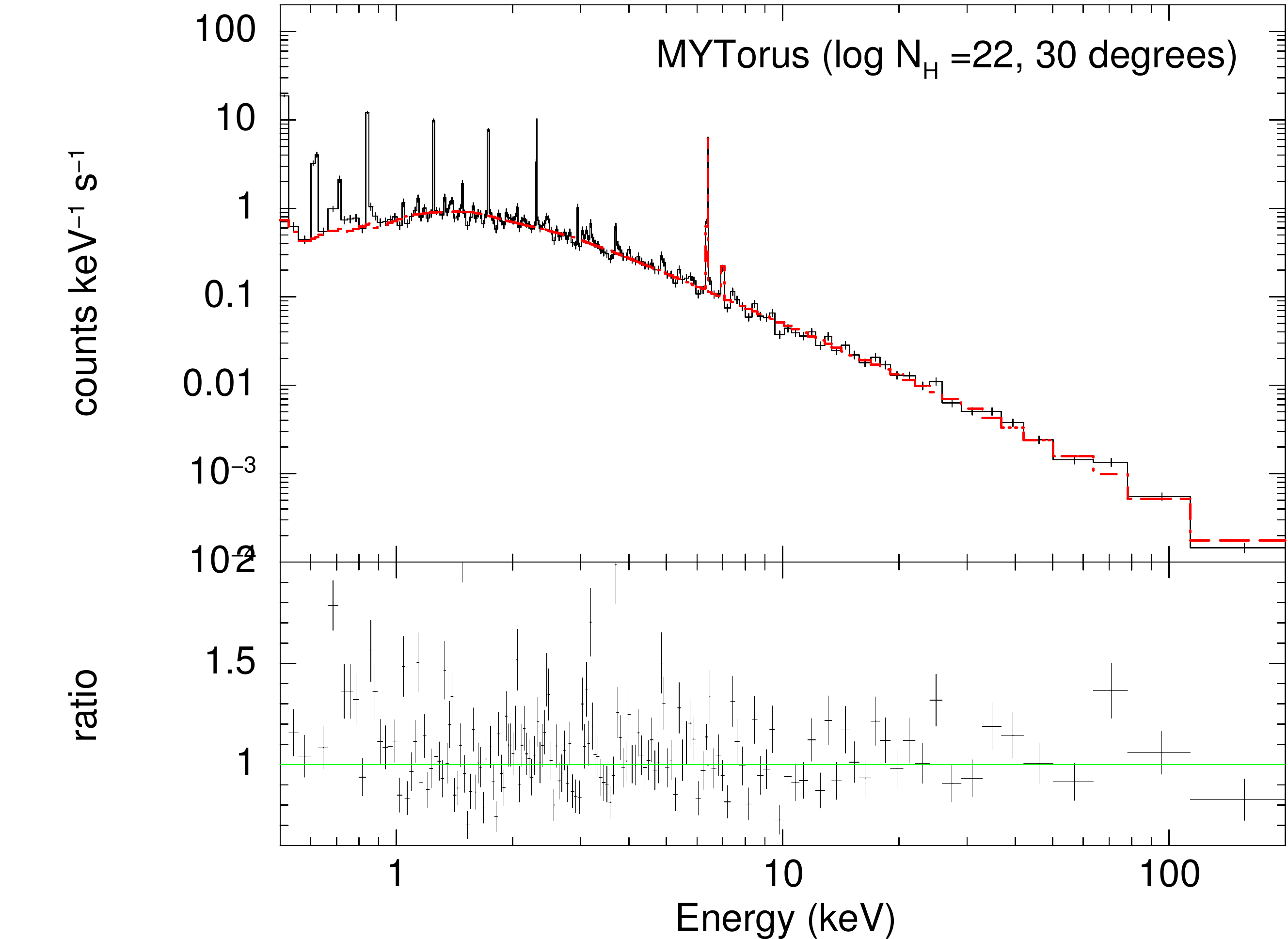}~
\includegraphics[height=8.8cm,angle=-90]{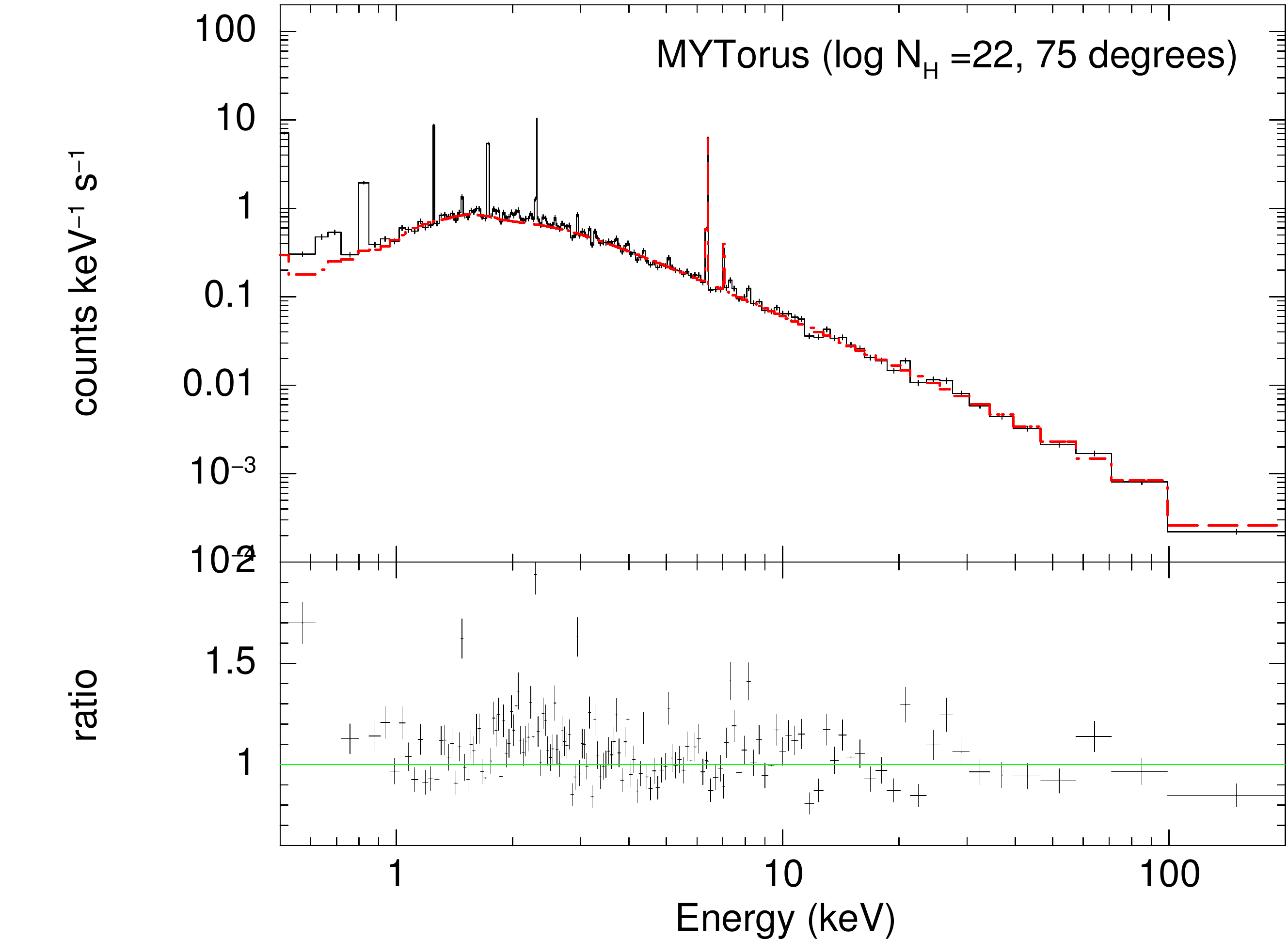}\\
\includegraphics[height=8.8cm,angle=-90]{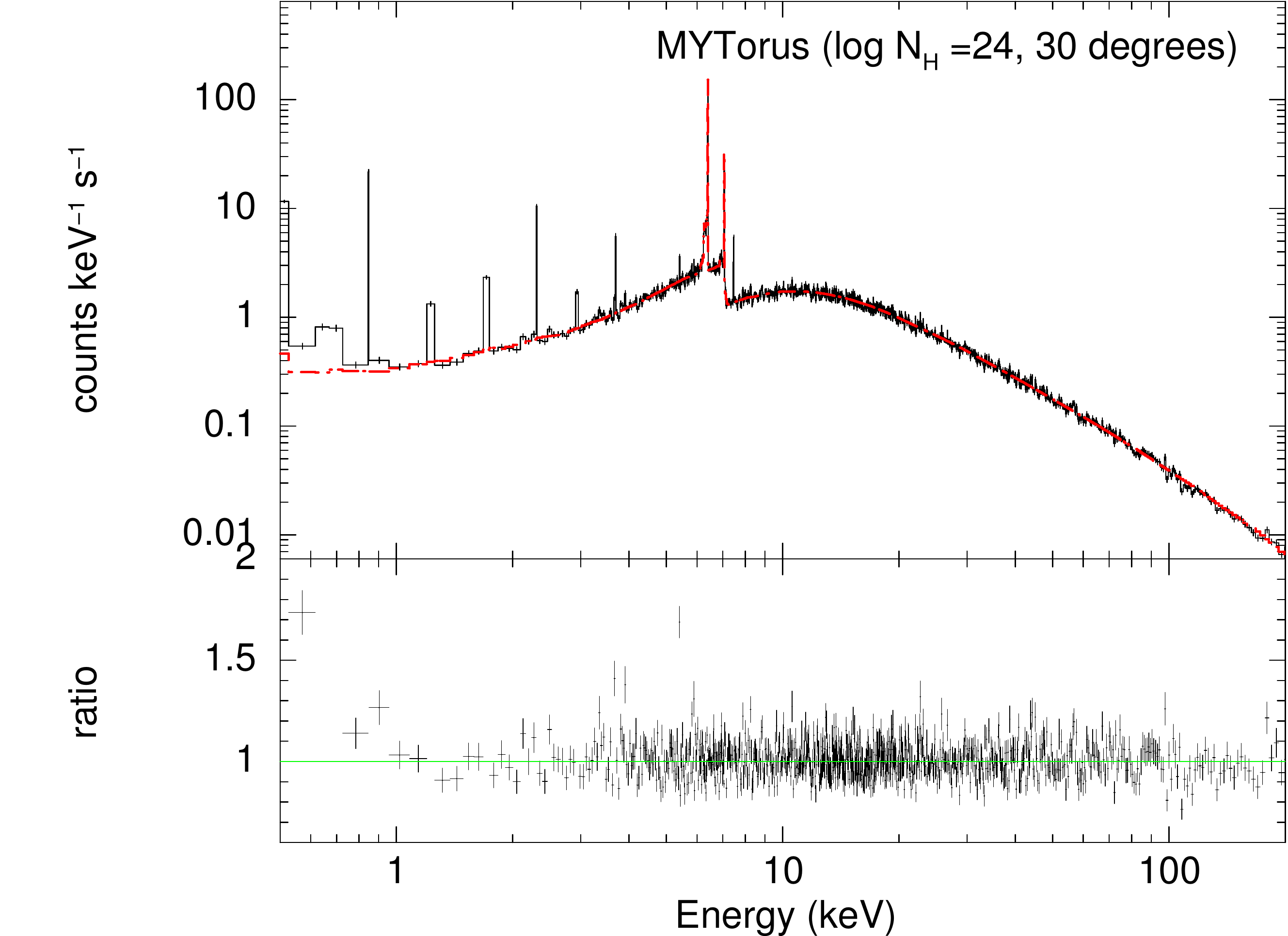}~
\includegraphics[height=8.8cm,angle=-90]{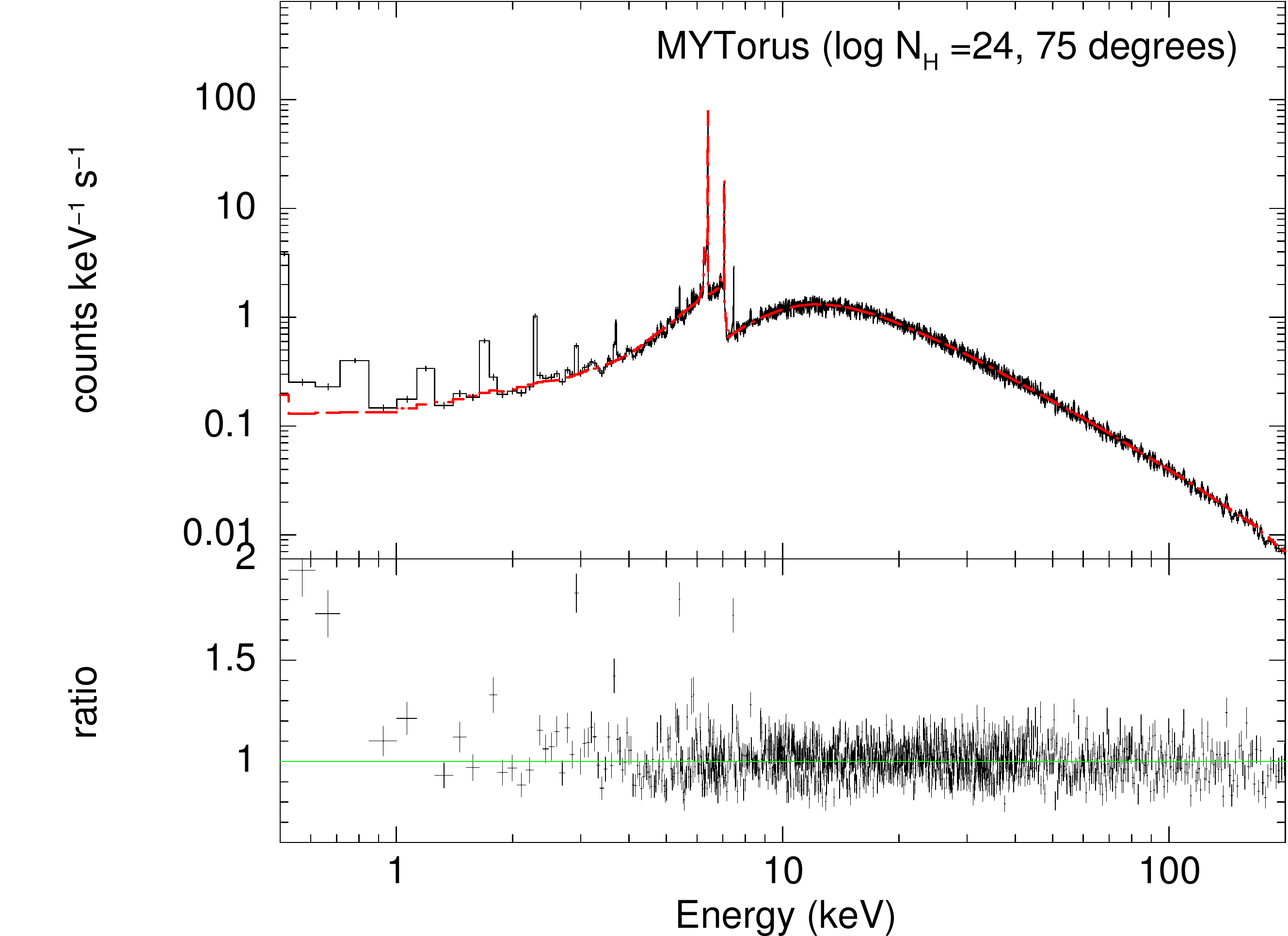}\\
\caption{\label{fig:mytorus}Comparison between \textsc{MYTorus} and our simulations assuming the geometry of Fig.\,\ref{ill:mytorus} for two viewing angles and two equatorial column densities. For both models we show only reprocessed radiation. Top: $i=30$\,deg; bottom: $i=75$\,deg. Left: $N_\mathrm{H,eq}=10^{22}$\,cm$^{-2}$; right: $N_\mathrm{H,eq}=10^{24}$\,cm$^{-2}$. In each panel, the top subplot shows the \textsc{MYTorus} model (black line) and our simulations (crosses); the bottom subplot shows the ratio between our simulations and the \textsc{MYTorus} model. In all cases, the primary continuum is defined as a power law between 1 and 500\,keV with a photon index $\Gamma=1.9$. Spectra are averaged over 2 deg in inclination. $10^{10}$ photons were generated.}
\end{figure*}

\textsc{MYTorus} \citep{Murphy2009} is a model for the reflection and transmission of X-rays based on Monte Carlo simulations, which is available in \textsc{xspec} in the form of model tables. \textsc{MYTorus} implements a toroidal geometry as shown on Fig.\ \ref{ill:mytorus} with a half-opening angle $\phi=60$\,deg, which corresponds to a torus minor radius $r$ (the radius of the tube) half the major radius $R$ (the distance from the center of the tube to the center of the torus). The central source is isotropic and the primary X-ray emission distribution follows a power law with a maximum energy. Because of the circular shape of the section of the torus, primary photons crossing the torus do not see the same hydrogen column densities, but rather
\begin{equation}
\label{eq:NH}
\left\{\begin{array}{ll}
N_\mathrm{H}(\vartheta)=N_\mathrm{H,eq}\left( 1- \frac{R^2}{r^2}  \cos^2 \vartheta\right)^{1/2} , &\cos \vartheta<\frac{r}{R},\\
N_\mathrm{H}(\vartheta)=0 , &\cos \vartheta>\frac{r}{R}\end{array}\right.,
\end{equation}
where $N_\mathrm{H,eq}$ is the equatorial column density, i.e.\ the column density seen by a photon that crosses the torus radially in the central plane of the torus, and $\vartheta$ is the angle between the direction of the primary photon and the perpendicular to the torus.

Different equatorial column densities can be modeled, from an optically thin torus with $N_\mathrm{H,eq}=10^{22}$\,cm$^{-2}$ to an optically thick torus with $N_\mathrm{H,eq}=10^{25}$\,cm$^{-2}$. The medium has an equivalent temperature of $10^4$\,K, meaning that H and He can be considered ionized and Fe ions up to Fe\,{\sc xvii} are present, i.e. there is no significant shift in the iron fluorescence line energy. Only Fe\,K$\alpha$ (single line representing Fe\,K$\alpha_1$+Fe\,K$\alpha_2$) and Fe\,K$\beta$ fluorescence lines are included. Apart from the geometry, the main difference between the \textsc{pexrav}/\textsc{pexmon} and \textsc{MYTorus} models is the fact that, in the latter case, the reflector is not necessarily optically thick, and therefore a transmitted component exists.

The physics included in \textsc{MYTorus} is practically identical to that included in the free-electron configuration of \name, and therefore constitutes an important benchmark for the validation of our code. Figure~\ref{fig:mytorus} shows the comparison between \textsc{MYTorus} and our simulations for different inclination angles ($i=30$\,deg, i.e.\ with a direct view of the central source and $i=75$\,deg, i.e.\ through the torus) and different equatorial column densities. For visual clarity, only the reprocessed X-ray radiation is shown. The primary continuum is defined as a power law  with a photon index $\Gamma=1.9$. The termination energy was set to 500\,eV, similar to the minimum energy in {\sc MYTorus}. The maximum energy was set to 500\,keV. \textsc{MYTorus} implements only the solar element abundances from \citet{Anders1989}. The {\sc MYTorus} and \name\ simulations match perfectly, and the only differences are attributed to the fluorescence lines of low-Z elements which are not included in {\sc MYTorus}.

\begin{figure}[tbp]
\includegraphics[width=8.8cm]{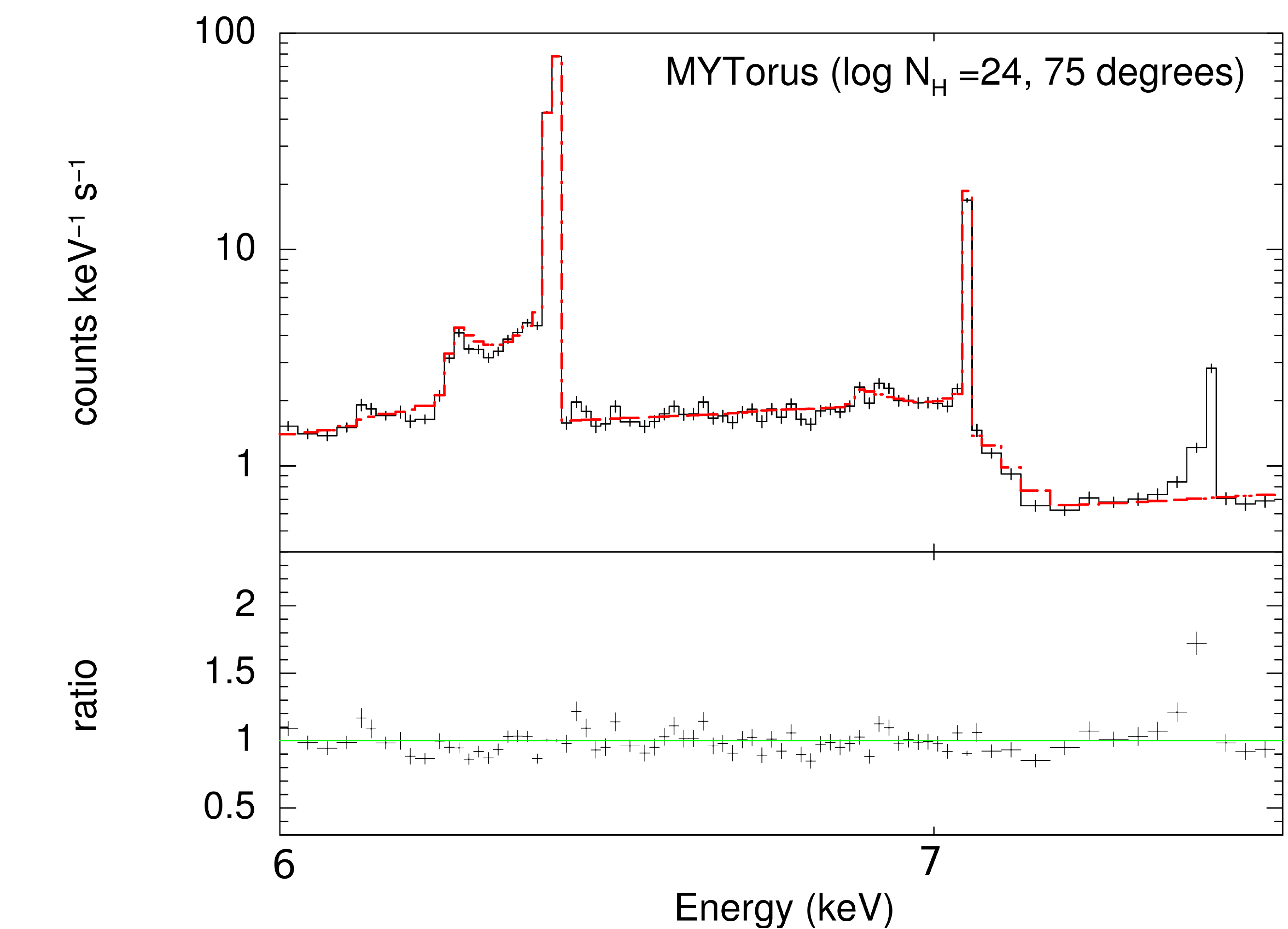}
\caption{\label{fig:mytoruszoom}Comparison between \textsc{MYTorus} and our simulations for $i=75$\,deg and $N_\mathrm{H,eq}=10^{24}$\,cm$^{-2}$, limited to the 6--7.5\,keV region. The \textsc{MYTorus} model is shown with a black line and our simulation is shown with a red line. The primary continuum is the same as in Fig.~\ref{fig:mytorus}.}
\end{figure}
Figure~\ref{fig:mytoruszoom} shows a zoom on the 6--7.5\,keV region for $i=30$\,deg. Again, the match is excellent. In particular, the shape of the Compton shoulder and the line intensities are perfectly reproduced. The only significant difference is the presence of fluorescence lines, such as Ni\,K$\alpha$ at E=7.470\,keV, which are not included in \textsc{MYTorus}.

The remarkable agreement at all column densities between the two simulations, which were developed completely independently, validates the implementation of the physical processes in \name. 

\begin{figure*}[tbp]
\includegraphics[height=8.8cm,angle=-90]{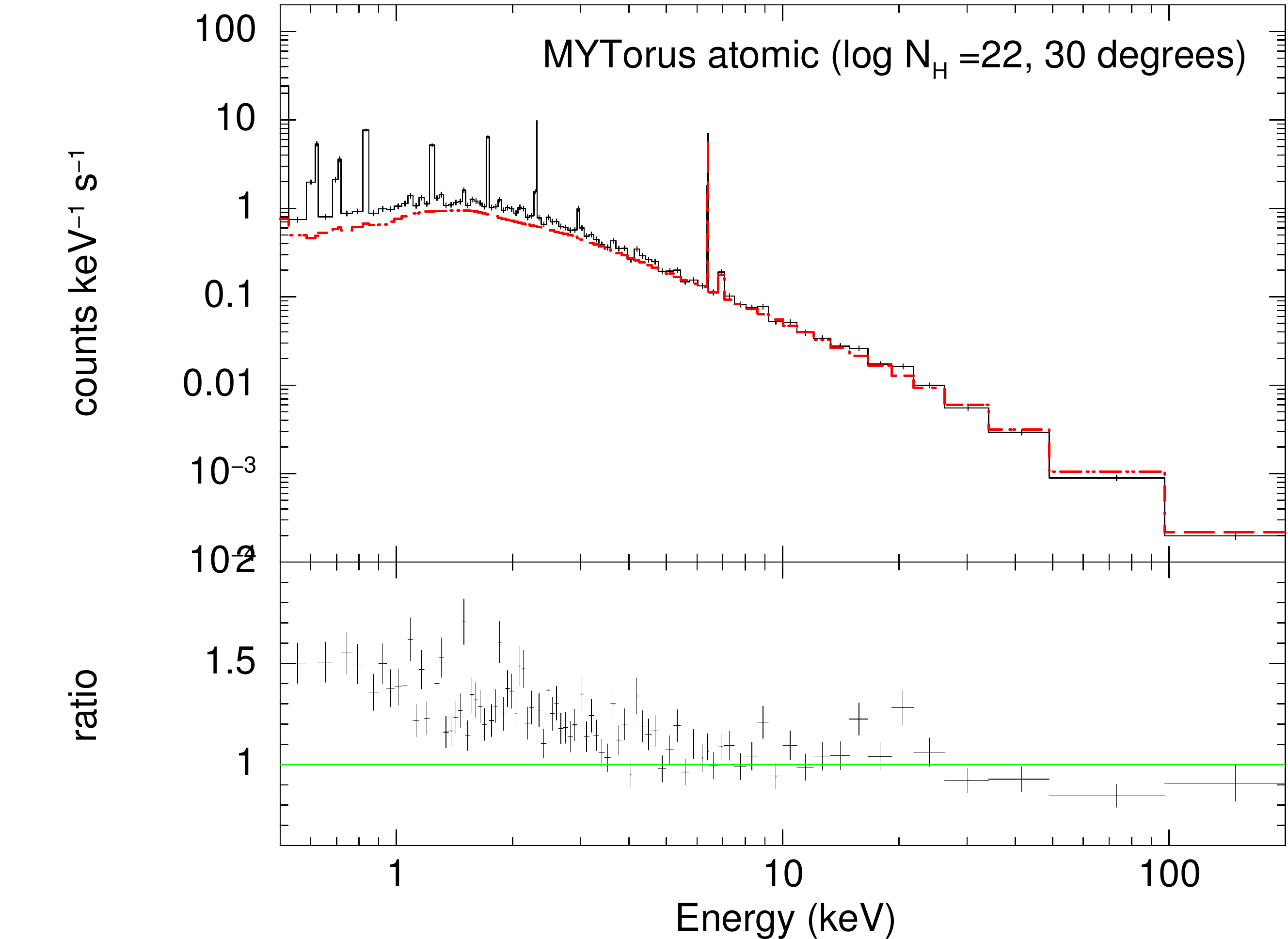}~
\includegraphics[height=8.8cm,angle=-90]{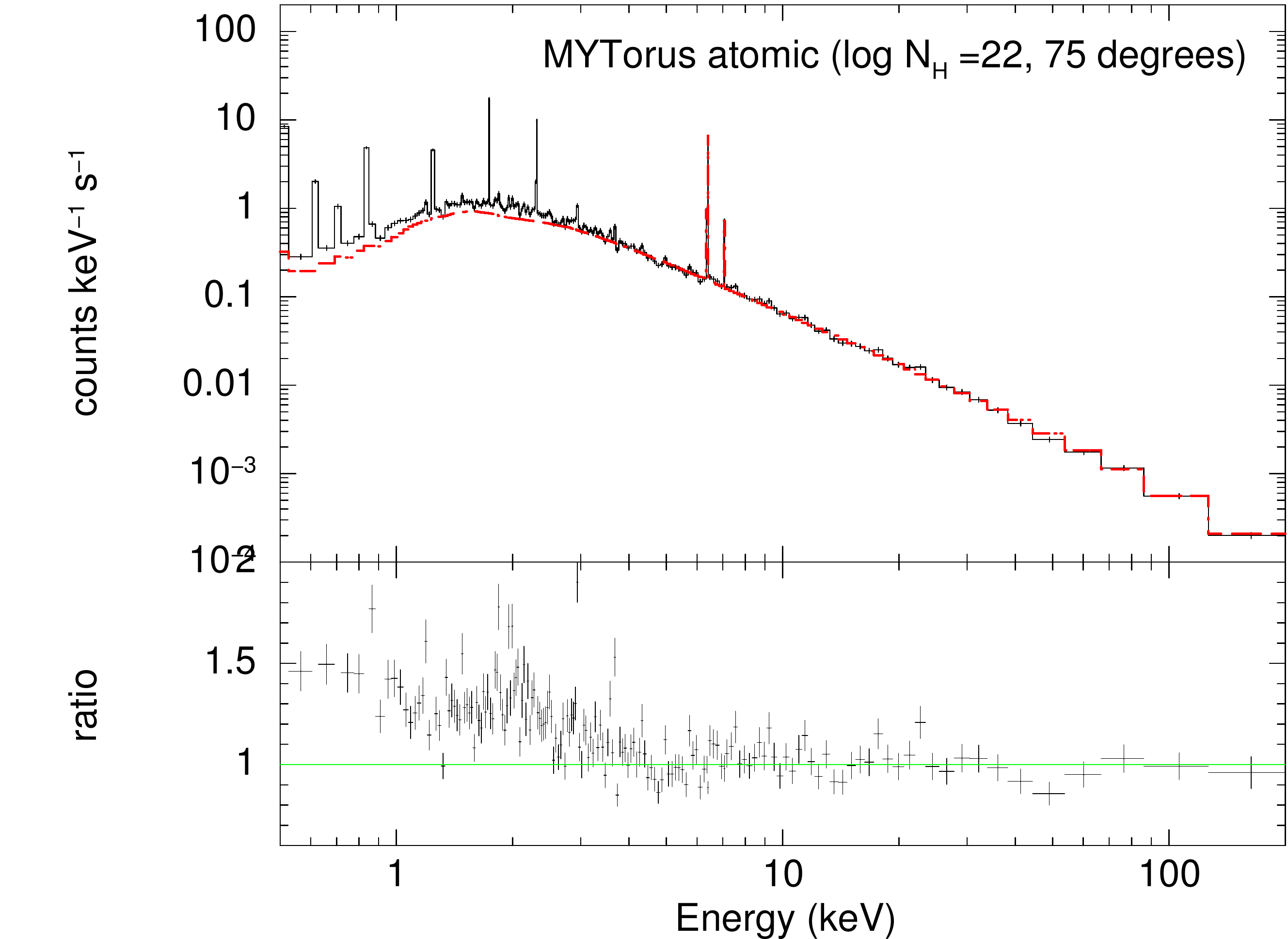}\\
\includegraphics[height=8.8cm,angle=-90]{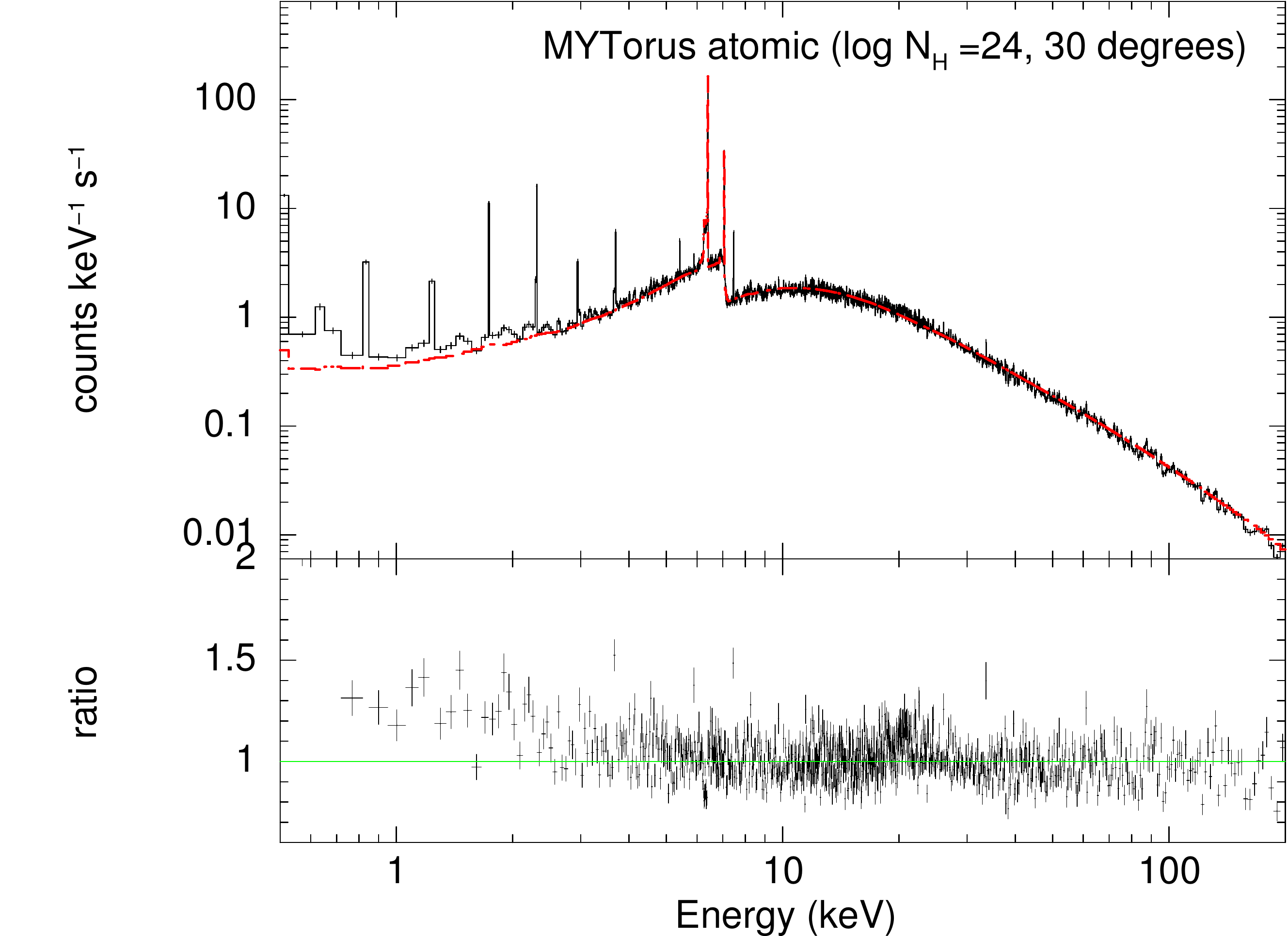}~
\includegraphics[height=8.8cm,angle=-90]{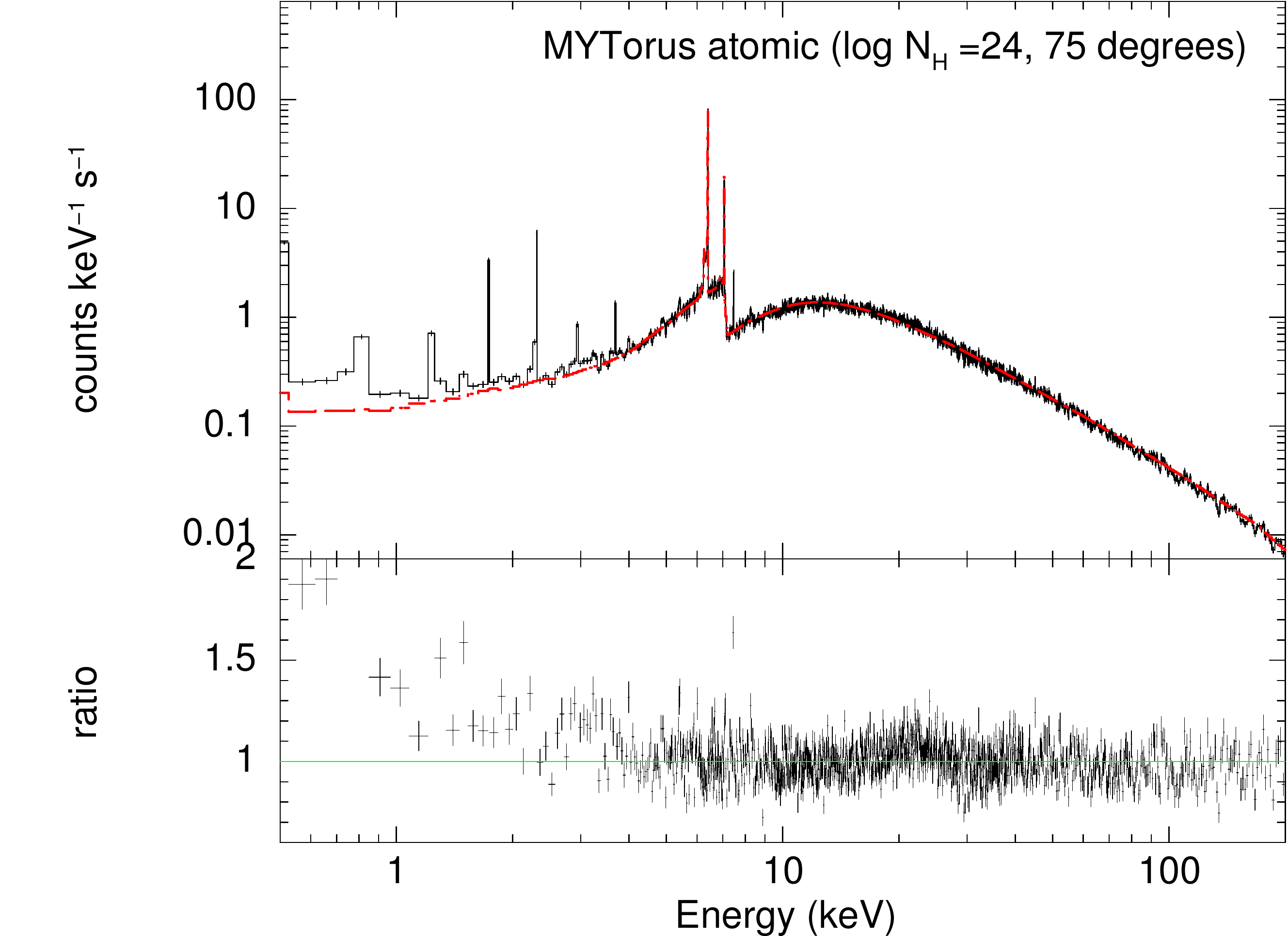}\\
\caption{\label{fig:mytorus_atomic}Comparison between \textsc{MYTorus} and our simulations assuming the geometry of Fig.\,\ref{ill:mytorus} for two viewing angles and two equatorial column densities taking into account the cross-sections for bound electrons. For both models we show only reprocessed radiation. Top: $i=30$\,deg; bottom: $i=75$\,deg. Left: $N_\mathrm{H,eq}=10^{22}$\,cm$^{-2}$; right: $N_\mathrm{H,eq}=10^{24}$\,cm$^{-2}$. In each panel, the top subplot shows the \textsc{MYTorus} model (black line) and our simulations (crosses); the bottom subplot shows the ratio between our simulations and the \textsc{pexmon} model. In all cases, the primary continuum is defined as a power law starting at 1\,keV with a photon index $\Gamma=1.9$. Spectra are averaged over 2 deg in inclination. $10^{10}$ photons were generated.}
\end{figure*}

Figure \ref{fig:mytorus_atomic} shows the comparison between \textsc{MYTorus} and \name\ for the same geometry, but using the atomic configuration. The source parameters are identical to those used in Fig.~\ref{fig:mytorus}. We observe an excess of scattered emission below $\sim 5$\,keV, which can exceed 50\% at 1\,keV in the  $N_\mathrm{H,eq}=10^{24}$\,cm$^{-2}$ case. This effect is due to the dominance of Rayleigh scattering at low energy. Such an effect has already been pointed out by \citet{Liu2014} (see Sect.~\ref{sec:geant4}). Another difference is that the peak of the reflection hump is located at slightly higher energy than in the free-electron configuration, which results in a bump in the ratio of the two models around 20\,keV. Finally, the shape of the Compton shoulder of the Fe\,K$\alpha$ fluorescence line is modified, which creates strong residuals below 6.4\,keV. Such behavior has also been observed by \citet{Furui2016} using the MONACO model (see Sect.~\ref{sec:monaco}).

\subsection{Spherical-toroidal geometry}
\label{sec:mb}

\begin{figure}[tbp]
\includegraphics[width=8.8cm]{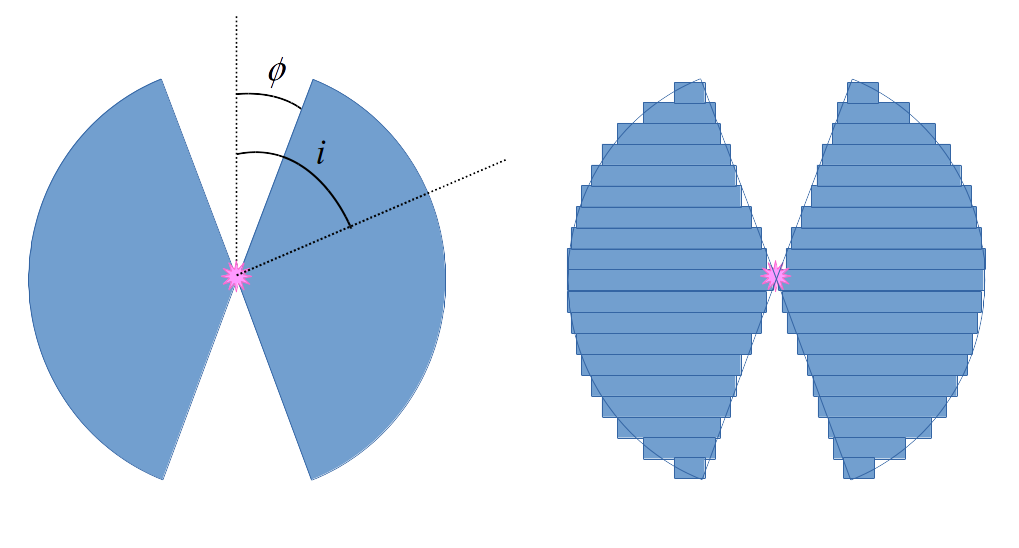}\\
\caption{\label{ill:mb}Spherical-toroidal geometry of the \textsc{BNTorus} model of \citet{Brightman2011} (Left). The reflector is a sphere with a conical opening of angle $\phi$, whose section is shown in blue. The source of X-ray photons is the magenta star in the center, which is isotropic. The inclination angle $i$ is defined as the angle between the normal to the  plane of the torus and the observer. In \name, we model this shape by superimposing a number of annuli (right) that approximate the spherical-toroidal geometry.}
\end{figure}

\citet{Brightman2011} further developed the \textsc{BNTorus} model from an original Monte Carlo code from \citet{Nandra1994} to simulate an approximately toroidal geometry based on a sphere with polar conical holes. The geometry is illustrated in Fig.~\ref{ill:mb} left. A similar geometry was used by \citep{Ikeda2009}, but their model was not publicly released, and therefore it cannot be used for comparison. In \name, we can approximate this geometry by superimposing annuli with outer radii approximating a sphere and inner radii approximating a cone of given aperture (Fig.~\ref{ill:mb} right). We also use this geometry to validate the propagation of photons in multiple objects in \name. For these simulations, the original geometry has been simulated using 51 annuli and the free-electron configuration has been used to match the assumptions of \citet{Brightman2011}.
 
\begin{figure}[tbp]
\includegraphics[height=8.8cm,angle=-90]{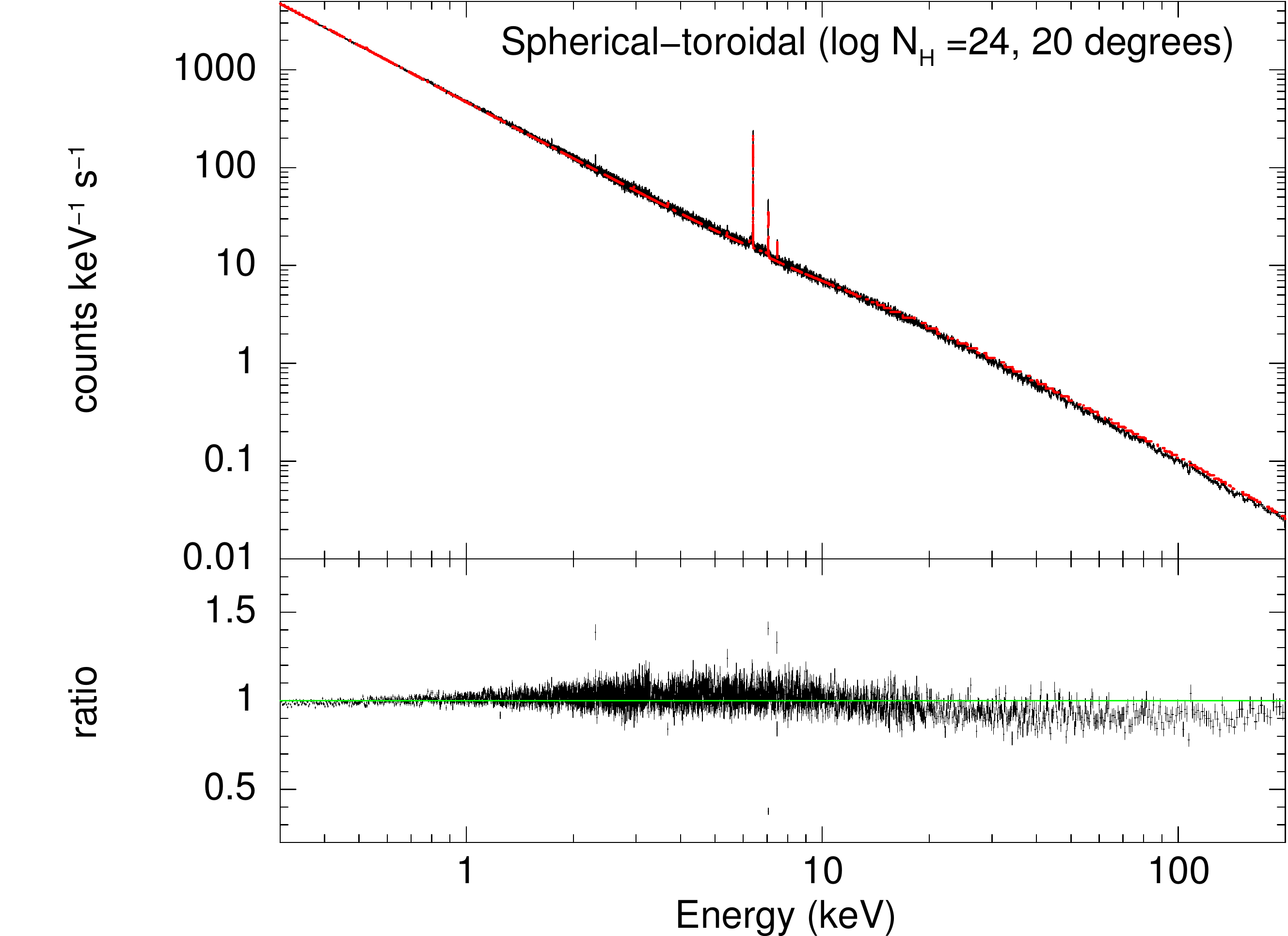}\\
\includegraphics[height=8.8cm,angle=-90]{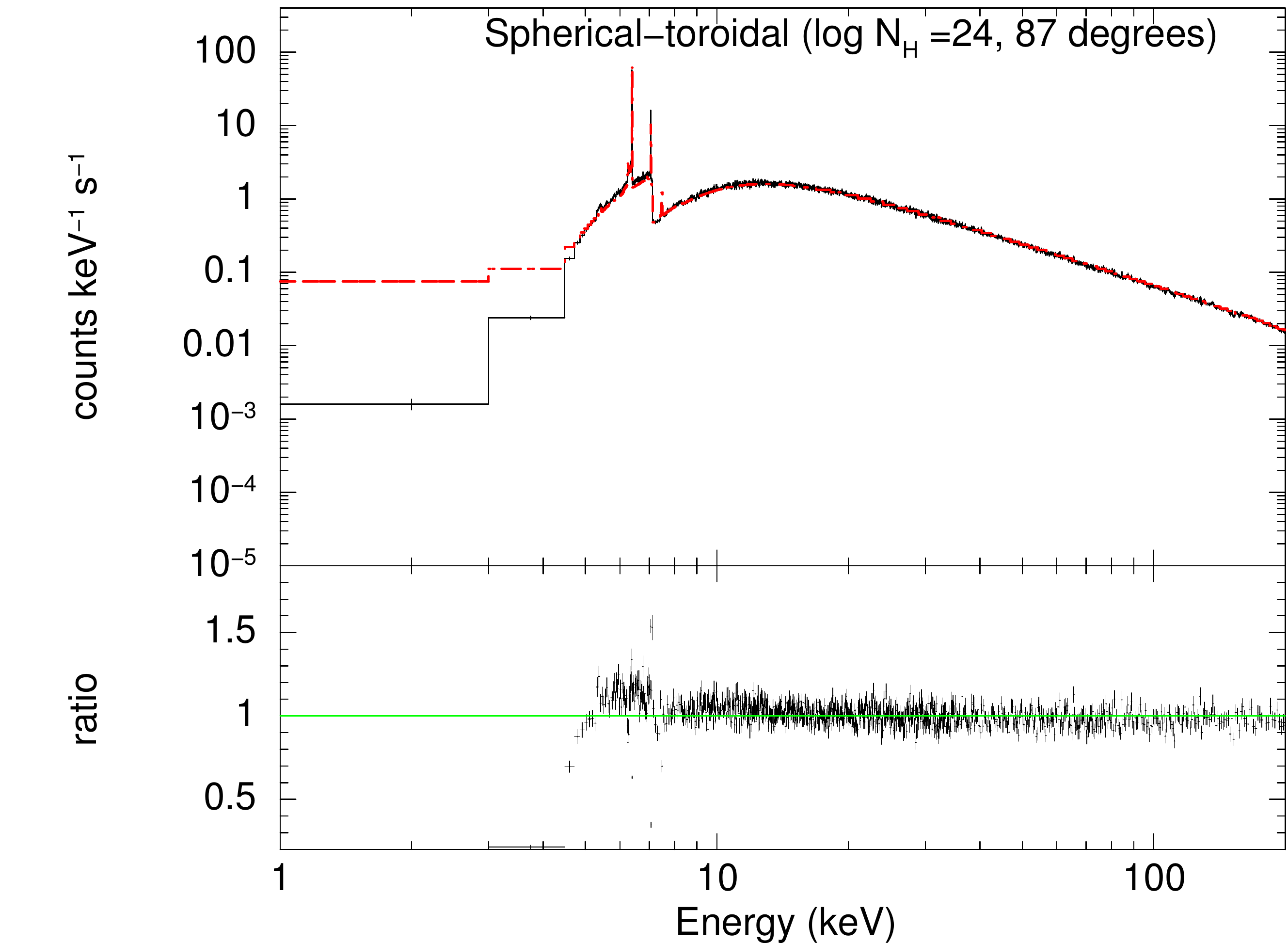}
\caption{\label{fig:mbdeg}Comparison between the spherical-toroidal geometry and our \name\ simulations assuming the \textsc{BNTorus} geometry (Fig.\,\ref{ill:mb}) for two viewing angles and an half-opening angle of $\theta_{\rm\,OA}=30$\,deg. Top: $i=20$\,deg; bottom: $i=87$\,deg. In each panel, the top subplot shows the spherical-toroidal model (solid line) and the \name\ simulation (crosses); the bottom subplot shows the ratio between our simulations and the spherical-toroidal model. In all cases, the primary continuum is defined as a power law starting at 1\,keV with an index $\Gamma=1.9$ up to 500\,keV. The equatorial column density is $N_\mathrm{H,eq}=10^{24}$\,cm$^{-2}$. The \textsc{BNTorus}  model has been approximated using 51 annuli in \name. Spectra are averaged over 2 deg in inclination. $10^{10}$ photons were generated.}
\end{figure}
Fig.\ref{fig:mbdeg} shows the comparison between the spherical-toroidal model of \citet{Brightman2011} and our \name\ simulations for two viewing angles. The primary continuum is defined as a power law  with an index $\Gamma=1.9$ up to 500\,keV. The termination energy has been set to 300\,eV. As for \textsc{MYTorus}, \textsc{BNTorus} uses the solar element abundances from \citet{Anders1989} and a physics similar to the free-electron configuration. The agreement is very good in the case of low inclination at all column densities. In the high-inclination case, while agreement is excellent above 10\,keV, there is a strong difference at low energy with the \textsc{BNTorus} model presenting a strong tail that is not present in \name. This difference becomes more and more important with increasing column densities. We thus confirm the results of \citet{Liu2015} based on Geant4 simulations (see Sect.~\ref{sec:geant4}), which identified a problem in the \citet{Brightman2011} code. \citet{Liu2015} interpret this behavior by postulating that the \textsc{BNTorus} code fails to treat correctly photons that are backscattered by the torus. Our simulations support this interpretation.

\subsection{Geant4-based models}
\label{sec:geant4}

Geant4\footnote{http://geant4.cern.ch} \citep{Agostinelli2003,Allison2006} is a C++ framework to simulate interactions of particles through matter. Geant4 has originally been developed in the context of high-energy physics to simulate the interactions of particles accelerated in particle accelerators inside detectors. As such, it implements a huge array of physical processes, for example, hadron physics, and very complex geometrical shapes to be able to construct any kind of detector. Although it is not the primary purpose of Geant4, electromagnetic processes are included. It is therefore straightforward to build a simulation tool similar to that presented in this work. Such an approach has been developed by \citet{Liu2014} and \citet{Liu2015}.

\begin{figure}[tbp]
\includegraphics[width=8.8cm]{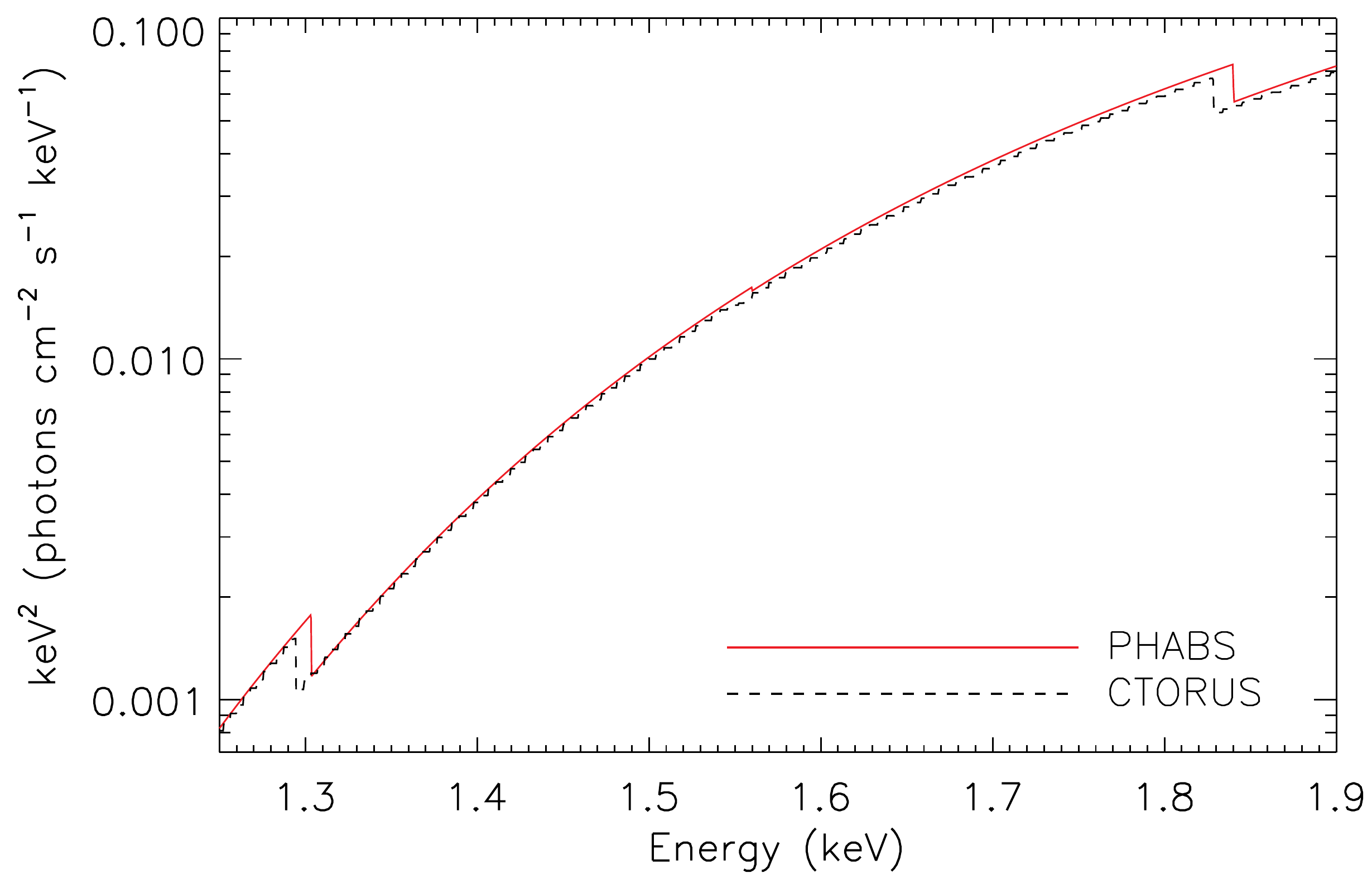}\\
\caption{\label{fig:epdl}Spectrum of an absorbed power law simulated with the Geant4-based model of \citet{Liu2014} compared with the \textsc{xspec} model \textsc{phabs}; $N_\mathrm{H}$ was set to $5\times 10^{22}$\,cm$^{-2}$. The figure shows that the absorption edges are clearly at different energies.}
\end{figure}

While extremely powerful, the Geant4 framework has a number of intrinsic limitations. For instance, the photoelectric cross-sections used by Geant4 have been derived from the EPDL89 library \citep{Cullen1990}, instead of the cross-sections from \citet{Verner1995} and \citet{Verner1996}, which are commonly used in astronomy; this leads to differences in the resulting spectrum which can be very significant in high-resolution spectra. Fig.~\ref{fig:epdl} shows the example of an absorbed power law, where the absorption edges are found at clearly different energies with differences of 20\,eV or more. It is in principle possible to replace the cross-sections, since they are defined in the low-energy electromagnetic process data pack (G4EMLOW), but it is very difficult to do in practice.

Likewise, fluorescence energies in G4EMLOW originate from the EPDL89 library, and do not match the energies measured in the laboratory by \citet{Bearden1967}. As an example, the Fe\,K$\alpha_1$ 6400\,eV fluorescence line appears at 6362.71\,eV, instead of 6403.84\,eV; this is a difference of more than 40\,eV. This can fortunately be more easily fixed, and one of us (SP) has actually implemented in Geant4 the energies measured by \citet{Bearden1967}, which can be selected by the user, starting with Version 10.1 (G4EMLOW version 6.41).

Geant4 implements the correction for bound electrons, so the model of \citet{Liu2014} should be compared with \name\ simulations using the atomic configuration. However, the only model made available by \citet{Liu2014} and \citet{Liu2015} is a clumpy-torus model, CTorus\footnote{https://heasarc.gsfc.nasa.gov/xanadu/xspec/models/Ctorus.html}; we defer the analysis of clumpiness of the torus with \name\ to another paper.

\subsection{MONACO}
\label{sec:monaco}
MONACO is a hybrid simulation tool developed by \citet{Odaka2011}. The computation of interactions is performed with Geant4, while the physical processes have been implemented by the authors. MONACO also uses EPDL97 for the atomic cross-sections, and therefore includes the bound-electron corrections.

\citet{Furui2016} have recently applied the MONACO simulation tool to the AGN torus in a \textsc{MYTorus}-like configuration. Comparing with the the \textsc{MYTorus} simulations, they found that the shape Compton shoulder is strongly affected by the correction for bound electrons. As in the model of \citet{Liu2014}, \citet{Furui2016} have found that the location of the edges differ in MONACO and in \textsc{MYTorus}, because of the different origin of the cross-sections (see figure 2 in \citealp{Furui2016}).

Since no model based on MONACO has been released, a proper comparison cannot be performed.

\section{The RXTorus xspec models}
\label{sec:rxtorus}

\subsection{The model}
As a first application of \name, we developed a new model, \textsc{RXTorus}, which extends the simulations in \textsc{MYTorus} adopting the same source and absorber geometries, but allowing different values of the $r/R$ opening ratios (see Fig.\,\ref{ill:mytorus}). This model is publicly released in the form of \textsc{xspec} tables\footnote{The \name\ models are available at http://www.astro.unige.ch/reflex}. 

We generated the \textsc{RXTorus} models on a grid with the different parameters. For each simulation, $5\cdot10^8$\,photons were generated. The planar symmetry of the problem was used to double the statistics. \textsc{RXTorus} considers a primary X-ray spectrum in the form of a power law with a photon index $\Gamma$ between 1.0 and 3.0; the maximum photon energy is set to 300\,keV. As in \textsc{MYTorus}, it is the maximum energy of the power law, and not a cutoff energy. The \textsc{RXTorus} torus equatorial column density $N_\mathrm{H,eq}$ covers the range from $10^{22}$\,cm$^{-2}$ up to $10^{25}$\,cm$^{-2}$. The ratio $r/R$ varies from 1 (i.e.\ a horn torus, a torus without central hole, or $\phi=0$ in Fig.\ \ref{ill:mytorus}) to 0.01. As a function of this ratio, several quantities can be readily obtained as follow:
\begin{itemize}
\item The line-of-sight column density for a given inclination $i$ is given by Eq.\,(\ref{eq:NH}), with $\vartheta=i$.
\item The opening angle $\phi$, i.e., the angle above which the line of sight crosses the torus is given by
\begin{equation}
\cos \phi=\frac{r}{R}
\end{equation}
\item The torus covering fraction, i.e. the fraction of the primary photons that enters the torus is
\begin{equation}
\label{eq:cfact}
C_\mathrm{f}=\frac{r}{R}
\end{equation}
\end{itemize}

\noindent Spectra were binned in inclination angle with a 3-degree step. In order to allow the inclination parameter to cover the full 0--90 deg range, the first and last bins are assumed to be centered on 0 and 90 deg, respectively. For compatibility with \textsc{MYTorus}, the composition is taken from \citet{Anders1989}, but we provide tables with different metallicities from 0.3 to 2.0 solar. Both free-electron and atomic configurations are available.

The \textsc{RXTorus} model is subdivided into several table components to allow a greater flexibility, i.e.,
\begin{itemize}
\item The transmitted continuum contains only photons that did not undergo any physical process.
\item The scattered continuum contains all photons that underwent one or more Compton scatterings, but no photoionization and no fluorescence.
\item The fluorescence emission contains all photons that underwent at least one photoionization and subsequent fluorescence re-emission, plus any number of Compton scatterings.
\end{itemize}
In addition, a table with the full emission, i.e., the sum of the transmitted continuum, the scattered continuum, and the fluorescence emission, is also provided.

While the SXS micro-calorimeter on board Hitomi has unfortunately been lost with the mission itself, such instruments will certainly play a huge role in the future of X-ray astronomy. Therefore, we provide the \textsc{RXTorus} model with a energy resolution of $\Delta E=2$\,eV in the 0.3--10\,keV energy band. Above 10\,keV, given the lack of narrow spectral features, we use a logarithmic step of $\Delta E/E=0.001$.

\subsection{Applications -- the case of NGC\,424}

We applied \textsc{RXTorus} to the broadband X-ray spectrum of \object{NGC 424}. This source was selected because, at a luminosity distance of $d_{\rm\,L}=45.6$\,Mpc, it is one of the closest CT AGN detected by Swift/BAT \citep{Ricci2015} and observed by NuSTAR. We analyzed the XMM-Newton EPIC spectrum of the source using an 8.2\,ks observation (ID 0002942301, PI M. Guainazzi). The source was observed by NuSTAR for 15.5\,ks (ID 60061007002) as a part of the campaign aimed at following up AGN detected by Swift/BAT.
We reduced XMM-Newton \citep[EPIC/PN, MOS1 and MOS2,][]{Jansen2001} and NuSTAR FPMA/FPMB \citep{Harrison2013} data following the standard guidelines \footnote{See \citet{Ricci2016} for details on the data reduction procedure adopted.}.

The spectral model we used consists in the \textsc{RXTorus} model, which takes into account absorbed continuum, scattered radiation, and fluorescent lines; two collisionally ionized plasmas (\textsc{apec}); and three Gaussian lines (\textsc{zgauss}). We took into account absorption by Galactic material using the \textsc{tabs} model \citep{Wilms2000}, fixing $N^{\rm\,Gal}_{\rm\,H}=1.8\times 10^{20}\rm\,cm^{-2}$ \citep{Kalberla2005}. To the models we also added a cross-calibration constant to take into account possible variability between the non-simultaneous XMM-Newton and NuSTAR observations. The addition of two \textsc{apec} components and three Gaussian lines significantly improved the fit.

\noindent We first applied the non-atomic version of \textsc{RXTorus}, i.e.,

\medskip
\noindent\textsc{tabs$_{\rm\,Gal}\times$(atable\{RXTorus.mod\} + 3$\times$zgauss + 2$\times$apec)}.
\medskip

\noindent This results in a good fit: $\chi^2$=203.3 for 205 degrees of freedom (DOF). The two thermal plasmas have temperatures of $kT_{1}=0.82^{+0.09}_{-0.08}$ and $kT_{2}=0.10^{+0.01}_{-0.01}$, while the lines have energies of $6.97^{+0.11}_{-0.10}$, $7.90^{+0.28}_{-0.12}$ and $5.35^{+0.11}_{-0.08}$. The equivalent widths of the lines are $235^{+110}_{-63}$\,eV, $573^{+118}_{-288}$\,eV, and $270^{+95}_{-82}$\,eV, respectively. In Figure\,\ref{ill:contRXtorus} we illustrate the contour plot of the inclination angle versus the ratio of the major to minor radius of the torus.

\noindent We then applied the atomic version of \textsc{RXTorus}, which also takes into account Rayleigh scattering, i.e.,

\medskip
\noindent\textsc{tabs$_{\rm\,Gal}\times$(atable\{RXTorus\_atom.mod\} + 3$\times$zgauss + 2$\times$apec)}.
\medskip

\noindent We find that this model can reproduce equally well the data ($\chi^2$/DOF=204.6/205).  The temperatures of the \textsc{apec} components obtained are consistent with those we found using the non-atomic version of \textsc{RXTorus}: $kT_{1}=0.82^{+0.09}_{-0.08}$ and $kT_{2}=0.10^{+0.01}_{-0.01}$. Similarly,  the three emission lines also yield consistent energies and equivalent widths: $6.99^{+0.12}_{-0.12}$ ($202^{+87}_{-107}$\,eV), $7.89^{+0.27}_{-0.12}$ ($592^{+237}_{-400}$\,eV) and $5.35^{+0.10}_{-0.07}$ ($290^{+86}_{-120}$\,eV).

The results of the fits obtained with these two models are reported in Table\,\ref{tab:resultsfit}. Both models find that the X-ray source of NGC\,424 is heavily obscured and has a very soft primary X-ray emission, which is in agreement with previous studies carried out using NuSTAR data (\citealp{Balokovic2014} and \citealp{Brightman2015}). The values of the inclination angle and of $\log R/r$ obtained with the two models are significantly different. The fit carried out with the atomic model results in a lower value of the inclination angle and of the ratio of the minor to major axis of the torus (i.e. a larger covering factor). The hydrogen column density along the line of sight is also different in the two cases, with values of $2.96^{+0.40}_{-0.51}\times10^{24}$\,cm$^{-2}$ and $1.45^{+0.36}_{-0.41}\times10^{24}$\,cm$^{-2}$ in the non-atomic and atomic configurations respectively. The column densities found using \textsc{RXTorus} are lower than the values inferred using \textsc{MYTorus} ($3\pm1\times10^{24}\rm\,cm^{-2}$ for the reflecting material and $>5\times 10^{24}\rm\,cm^{-2}$ for the primary X-ray emission; \citealp{Balokovic2014}) and \textsc{BNTorus} ($5.4^{+15.7}_{-1.2}$; \citealp{Brightman2015}). The covering factors of the torus obtained, calculated using Eq.\,\ref{eq:cfact} are $0.59^{+0.03}_{-0.04}$ and $0.79^{+NC}_{-0.02}$ for the non-atomic and atomic model, respectively. These values are larger than those obtained using \textsc{BNTorus} by \cite{Brightman2015}, who found a covering factor in the range $0.20-0.45$. 

\begin{table*}
\begin{center}
\caption[]{Results obtained by fitting the XMM-Newton and NuSTAR spectra of NGC\,424 with the non-atomic and atomic version of \textsc{RXTorus}. The table reports (2) the photon index, (3) the ratio between the major and minor torus axis, (4) the inclination angle, (5) the equatorial hydrogen column density in units of $10^{24}$\,cm$^{-2}$, and (6) the $\chi^2$ and degrees of freedom (DOF) of the fit. NC indicates an unconstrained upper or lower limit.}
\label{tab:resultsfit}
\begin{tabular}{cccccc}
\hline \hline\noalign{\smallskip}
(1) & (2) & (3) & (4) & (5) & (6)\\
\noalign{\smallskip}
& $\Gamma$ & $\log R/r$ & $i$ & $N_\mathrm{H,eq}$ & $\chi^2$/DOF  \\
\noalign{\smallskip}
\hline
\noalign{\smallskip}
Non-atomic & $2.49^{+0.16}_{-0.22}$   & $0.23^{+0.03}_{-0.02}$  & $62.48^{+0.01}_{-0.02}$  & $4.77^{+1.43}_{-1.08}$  &  203.3/205 \\
\noalign{\smallskip}
\noalign{\smallskip}
Atomic & $2.49^{+0.13}_{-0.10}$  & $0.10^{+0.01}_{-NC}$  & $34.5\phantom{0}^{+1.5\phantom{0}}_{-2.1\phantom{0}}$  & $5.33^{+0.67}_{-0.80}$ & 204.6/205   \\
\noalign{\smallskip}
\hline
\noalign{\smallskip}
\end{tabular}
\end{center}
\end{table*}

\begin{figure*}[tbp]
\includegraphics[width=8.8cm]{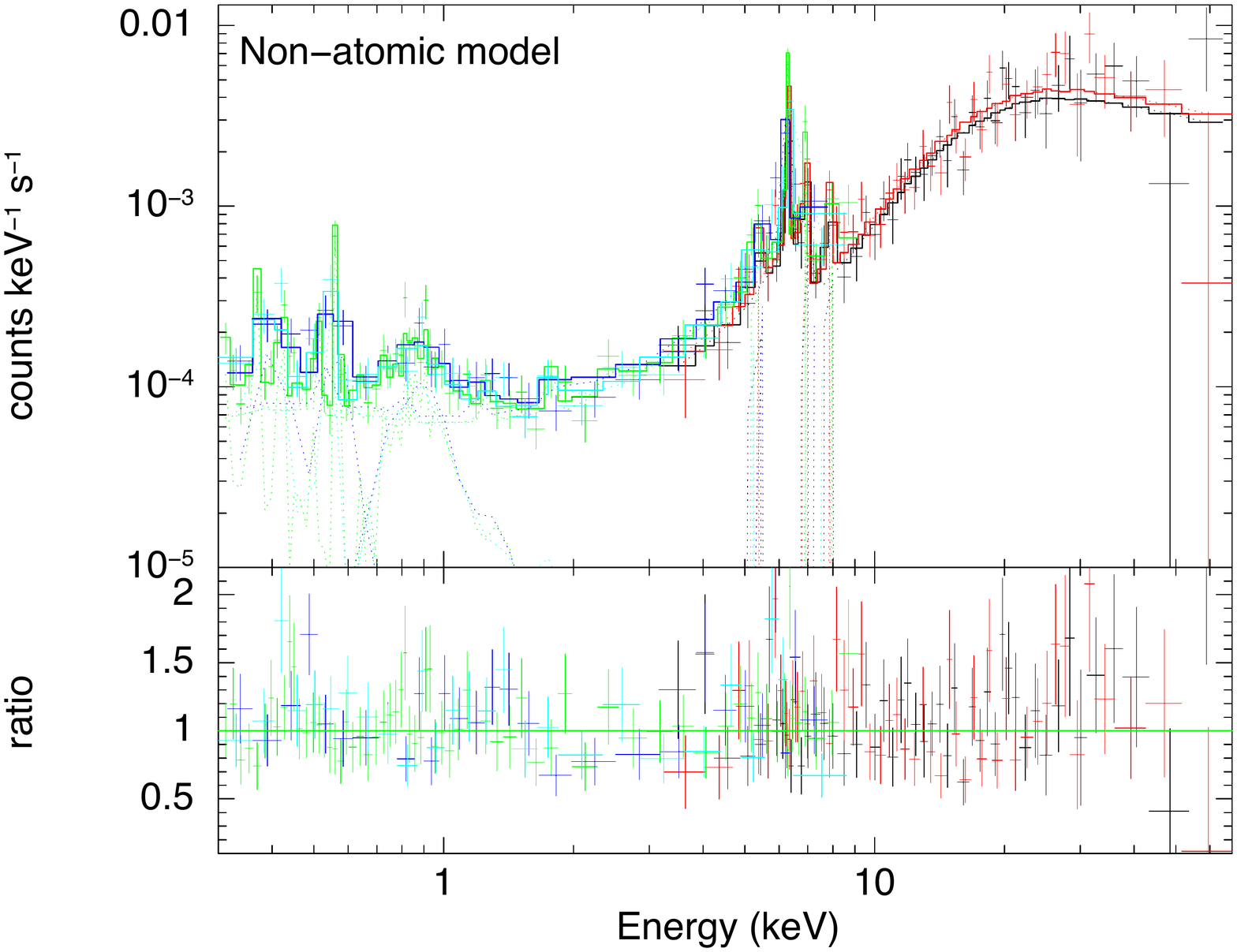}~
\includegraphics[width=8.8cm]{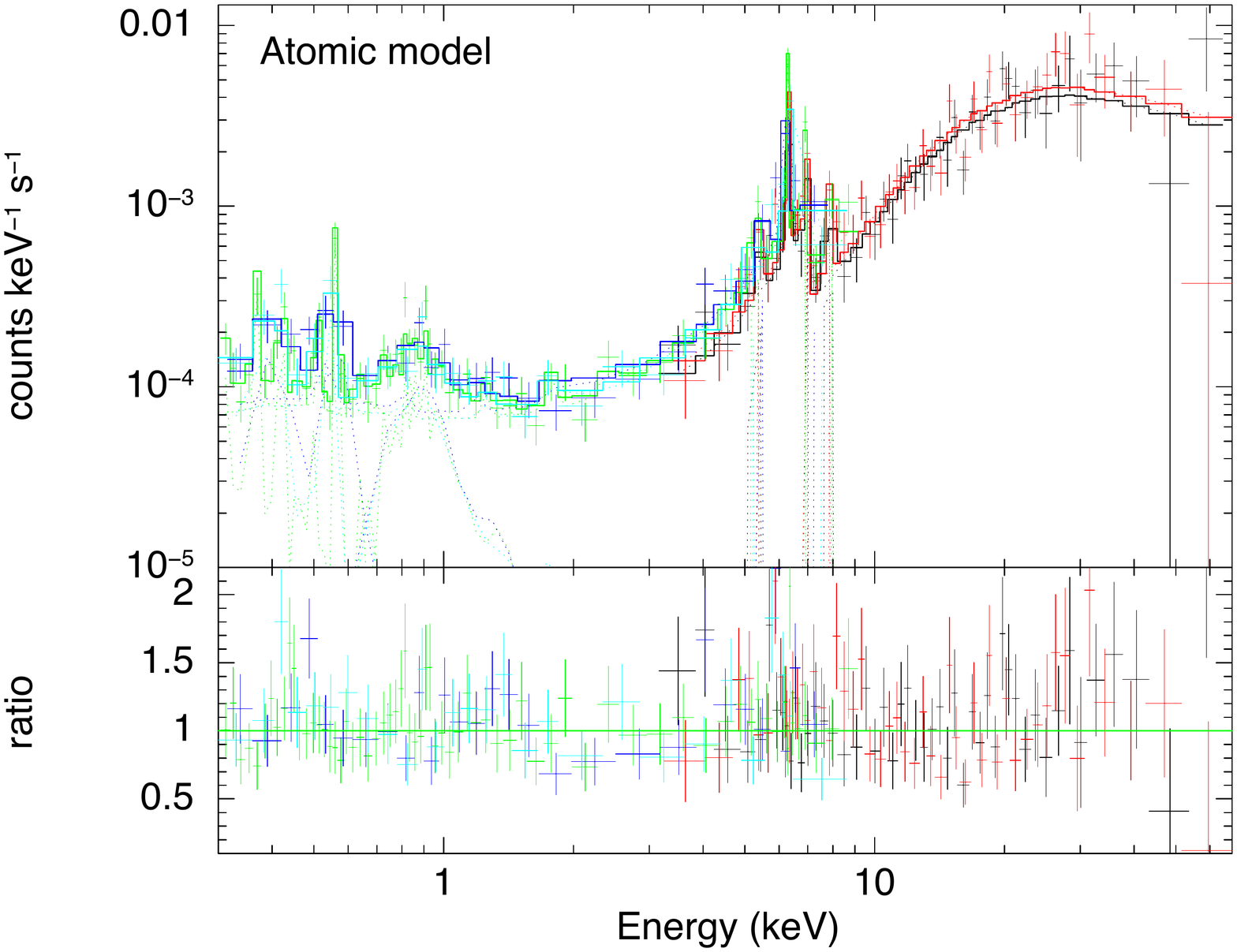}\\
\caption{\label{fig:NGC424spec} NuSTAR and XMM-Newton spectrum of NGC\,424 fitted with the non-atomic ({\it left panel}) and atomic ({\it right panel}) version of \textsc{RXTorus}. The spectral parameters obtained are reported in Table\,\ref{tab:resultsfit}.}
\end{figure*}

\begin{figure}[tbp]
\includegraphics[width=8.8cm]{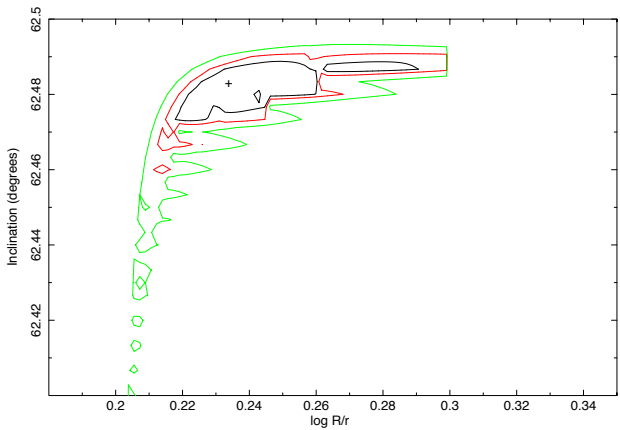}\\
\caption{\label{ill:contRXtorus} Contour plot of the inclination angle versus the ratio of the major to minor radius of the torus for the fit to the XMM-Newton and NuSTAR spectrum of NGC\,424 carried out with the non-atomic \textsc{RXTorus} model. The cross indicates the best fit. The black, red and green lines show the $\Delta\chi^2=$ 2,3, 4.61 and 9.21 contours respectively (corresponding to 68, 90 and 99\% respectively)}
\end{figure}

\section{Conclusions}
We have presented here \name, a new framework to self-consistently simulate absorption and reprocessed X-ray radiation in AGN for complex geometries, which we make publicly available. We showed the reliability of our approach by comparing the results of our simulations to existing X-ray spectral models. Contrarily to what has been commonly carried out in other models, \name\ implements two physical situations, namely the free-electron configuration where hydrogen and helium are completely ionized and all other electrons are ignored, and the atomic configuration where all atoms are neutral and Rayleigh scattering dominates over Compton scattering at low energies. The former configuration is used in \textsc{pexmon} \citep{Nandra2007}, \textsc{MYTorus} \citep{Murphy2009}, and \textsc{BNTorus} \citep{Brightman2011}, while the latter is consistent with Geant4-based models \citet{Liu2014} and MONACO \citep{Odaka2011}. Slight differences are observed with \textsc{pexmon}, which are probably due to the simplified treatment in this model. On the other hand, the \textsc{MYTorus} and \textsc{BNTorus} can be perfectly reproduced, with the exception that all fluorescent lines are included in \name, and that \textsc{BNTorus} is affected by a bug that overpredicts the transmission at low energy, which is an effect already noted by \citet{Liu2015}. Unfortunately, we cannot make a detailed comparison with either the \citet{Liu2014} model or MONACO, but we point out that the cross-sections used in these models differ from those used in the other models, which are taken from \citet{Verner1995} and \citet{Verner1996}; in particular edge energies are found to differ by more than 20\,eV in some cases, which will be easily observed by X-ray micro-calorimeters such as the one on a possible successor to Hitomi or the Athena X-IFU \citep{Barret2016}.

The issue of the choice of the physical configuration is not straightforward, since we expect that hydrogen and helium will be ionized in an accretion disk, but that the distant molecular torus is essentially neutral. \name\ offers the possibility to select the most suitable configuration for each object, allowing for instance the simulation of reflection on a warm accretion disk with free electrons surrounded by a cold molecular torus with bound electrons.

We introduce here the \textsc{RXTorus} model, which we release publicly. \textsc{RXTorus} is similar to the \textsc{MYTorus} \citep{Murphy2009} geometry, except that it allows a varying covering factor of the torus. In addition, both the free-electron and the atomic configurations can be selected. The model is distributed in three tables, i.e., absorbed continuum, scattered emission, and fluorescent lines, to allow greater flexibility. This is the first torus model with a varying covering factor that can be decomposed in several components. Our calculations are performed with a spectral resolution of 2\,eV below 10\,keV and allow us to use different abundances of the circumnuclear material. We applied this new model to reproduce the XMM-Newton and NuSTAR spectra of the CT AGN NGC\,424, and discussed the results obtained with or without considering Rayleigh scattering.

In the future, we will release new models corresponding to different geometrical situations, such as, for example, a clumpy torus model.

\begin{acknowledgements}

CR acknowledges financial support from FONDECYT 1141218, Basal-CATA PFB--06/2007 and from the China-CONICYT fund.

\end{acknowledgements}

\bibliographystyle{aa}
\bibliography{biblio}

\appendix

\section{K-shell (1s1/2) Fluorescence lines implemented in \name}
\label{sec:fluotable}

List of K-shell (1s1/2) fluorescence lines implemented in \name. The transition name is provided using Siegbahn notation. Energy is as determined by Bearden (1967), or, when absent, from the LBNL X-ray Data Booklet (Thompson et al., 2009). Yields (transition probabilities) are always taken from the LBNL X-ray Data Booklet.

\begin{tabular}{llrl}
\hline
\hline
\rule{0pt}{1.2em}Element& \multicolumn{1}{c}{Transition}& \multicolumn{1}{c}{Energy} & Yield\\
&&\multicolumn{1}{c}{eV}\\
\hline
\rule{0pt}{1.2em}%
C
&$\alpha_2KL_{II}$&277&	     0.000561488\\
&$\alpha_1KL_{III}$&277&	     0.0011206\\[2mm]
N
&$\alpha_2KL_{II}$&392.4&	     0.0010942\\
&$\alpha_1KL_{III}$&392.4&	     0.00218181\\[2mm]
O
&$\alpha_2KL_{II}$&524.9&	     0.00190768\\
&$\alpha_1KL_{III}$&524.9&	     0.00380027\\[2mm]
F
&$\alpha_2KL_{II}$&676.8&	     0.00306841\\
&$\alpha_1KL_{III}$&676.8&	     0.00610743\\[2mm]
Ne
&$\alpha_2KL_{II}$&848.6&	     0.00464329\\
&$\alpha_1KL_{III}$&848.6&	     0.00922967\\[2mm]
Na
&$\alpha_2KL_{II}$&1041.0	&     0.00668996\\
&$\alpha_1KL_{III}$&1041.0&	     0.0132959\\[2mm]
Mg
&$\alpha_2KL_{II}$&1253.60 &     0.00927327\\
&$\alpha_1KL_{III}$&1253.60 &     0.0184189\\[2mm]
Al
&$\alpha_2KL_{II}$&1486.27  &    0.0123699\\
&$\alpha_1KL_{III}$&1486.70  &    0.0245528\\
&$\beta_3KM_{II}$&1557.4	  &   0.0000755854\\
&$\beta_KM_{III}$&1557.4	  &   0.000150039\\[2mm]
Si
&$\alpha_2KL_{II}$&1739.38  &    0.0159791\\
&$\alpha_1KL_{III}$&1739.98  &    0.0317052\\
&$\beta_3KM_{II}$&1835.9	   &  0.000272402\\
&$\beta_KM_{III}$&1835.9	  &   0.000540444\\[2mm]
P
&$\alpha_2KL_{II}$&2012.7   &    0.020122\\
&$\alpha_1KL_{III}$&2013.7   &    0.0398749\\
&$\beta_3KM_{II}$&2139.0   &    0.000621469\\
&$\beta_KM_{III}$&2139.0   &    0.0012321\\[2mm]
S
&$\alpha_2KL_{II}$&2306.64  &    0.0247822\\
&$\alpha_1KL_{III}$&2307.84  &    0.0490644\\
&$\beta_3KM_{II}$&2468.1   &    0.00115591\\
&$\beta_KM_{III}$&2464.0   &    0.00228902\\[2mm] 
Cl
&$\alpha_2KL_{II}$&2620.78   &   0.0299473\\
&$\alpha_1KL_{III}$&2622.39   &   0.0592357\\
&$\beta_3KM_{II}$&2792.48   &   0.00190852\\
&$\beta_KM_{III}$&2792.61  &    0.00377764\\[2mm]
Ar
&$\alpha_2KL_{II}$&2955.63  &    0.0355868\\
&$\alpha_1KL_{III}$&2957.70  &    0.0703336\\
&$\beta_3KM_{II}$&3190.5   &    0.00291258\\
&$\beta_KM_{III}$&3190.5   &    0.00575646\\[2mm]
K
&$\alpha_2KL_{II}$&3311.1  &     0.0418599\\
&$\alpha_1KL_{III}$&3313.8  &     0.0826518\\
&$\beta_3KM_{II}$&3589.6   &    0.00399739\\
&$\beta_KM_{III}$&3589.6  &     0.00791008\\[2mm]
\hline
\end{tabular}

\begin{tabular}{llrl}
\hline
\hline
\rule{0pt}{1.2em}Element& \multicolumn{1}{c}{Transition}& \multicolumn{1}{c}{Energy} & Yield\\
&&\multicolumn{1}{c}{eV}\\
\hline
\rule{0pt}{1.2em}%
Ca
&$\alpha_2KL_{II}$&3688.09  &    0.0487196\\
&$\alpha_1KL_{III}$&3691.68  &    0.0961323\\
&$\beta_3KM_{II}$&4012.7   &    0.00517776\\
&$\beta_KM_{III}$&4012.7   &    0.0102489\\[2mm]
Sc
&$\alpha_2KL_{II}$&4086.1  &     0.0563841\\
&$\alpha_1KL_{III}$&4090.6   &    0.11113\\
&$\beta_3KM_{II}$&4460.5  &     0.00622301\\
&$\beta_KM_{III}$&4460.5   &    0.012304\\
&$\beta_2KN_{II}$&4486.5   &    0.00000019612\\
&$\beta_2KN_{III}$&4486.5   &    0.000000289041\\[2mm]
Ti
&$\alpha_2KL_{II}$&4504.86  &    0.0645982\\
&$\alpha_1KL_{III}$&4510.84  &    0.12714\\
&$\beta_3KM_{II}$&4931.81  &    0.00732672\\
&$\beta_KM_{III}$&4931.81  &    0.014475\\
&$\beta_2KN_{II}$&4962.3   &    0.000000590112\\
&$\beta_2KN_{III}$&4962.3   &    0.000000868503\\[2mm]
V
&$\alpha_2KL_{II}$&4944.64  &    0.0732827\\
&$\alpha_1KL_{III}$&4952.20  &    0.144051\\
&$\beta_3KM_{II}$&5427.29  &    0.00848408\\
&$\beta_KM_{III}$&5427.29  &    0.0167441\\
&$\beta_2KN_{II}$&5462.9   &    0.00000126461\\
&$\beta_2KN_{III}$&5462.9   &    0.00000185722\\[2mm]
Cr
&$\alpha_2KL_{II}$&5405.51  &    0.0825759\\
&$\alpha_1KL_{III}$&5414.72 &     0.16209\\
&$\beta_3KM_{II}$&5946.71  &    0.00949389\\
&$\beta_KM_{III}$&5946.71  &    0.018701\\
&$\beta_2KN_{II}$&5986.9	 &    2.5254e-06\\
&$\beta_2KN_{III}$&5986.9	 &    3.6955e-06\\[2mm]
Mn
&$\alpha_2KL_{II}$&5887.65  &    0.0917726\\
&$\alpha_1KL_{III}$&5898.75  &    0.179949\\
&$\beta_3KM_{II}$&6490.45  &    0.010936\\
&$\beta_KM_{III}$&6490.45  &    0.0215229\\
&$\beta_2KN_{II}$&6535.2   &    0.00000385518\\
&$\beta_2KN_{III}$&6535.2   &    0.00000564148\\[2mm]
Fe
&$\alpha_2KL_{II}$&6390.84  &    0.101391\\
&$\alpha_1KL_{III}$&6403.84  &    0.198621\\
&$\beta_3KM_{II}$&7057.98  &    0.0122111\\
&$\beta_KM_{III}$&7057.98  &    0.0240042\\
&$\beta_2KN_{II}$&7108.1	  &   0.00000600794\\
&$\beta_2KN_{III}$&7108.1	  &   0.00000877226\\[2mm]
Co
&$\alpha_2KL_{II}$&6915.30  &    0.11122\\
&$\alpha_1KL_{III}$&6930.32  &    0.21747\\
&$\beta_3KM_{II}$&7649.43  &    0.013502\\
&$\beta_KM_{III}$&7649.43  &    0.026513\\
&$\beta_2KN_{II}$&7705.9	  &   0.00000890651\\
&$\beta_2KN_{III}$&7705.9	  &   0.000012974\\[2mm]
Ni
&$\alpha_2KL_{II}$&7460.89  &    0.12106\\
&$\alpha_1KL_{III}$&7478.15  &    0.236419\\
&$\beta_3KM_{II}$&8264.66  &    0.014805\\
&$\beta_KM_{III}$&8264.66  &    0.0290299\\
&$\beta_2KN_{II}$&8328.6	  &   0.000012691\\
&$\beta_2KN_{III}$&8328.6	  &   0.000018445\\[2mm]
\hline
\end{tabular}

\begin{tabular}{llrl}
\hline
\hline
\rule{0pt}{1.2em}Element& \multicolumn{1}{c}{Transition}& \multicolumn{1}{c}{Energy} & Yield\\
&&\multicolumn{1}{c}{eV}\\
\hline
\rule{0pt}{1.2em}%
Cu
&$\alpha_2KL_{II}$&8027.83  &    0.131119\\
&$\alpha_1KL_{III}$&8047.78  &    0.255668\\
&$\beta_3KM_{II}$&8902.9   &    0.0158899\\
&$\beta_KM_{III}$&8905.29  &    0.0310908\\
&$\beta_2KN_{II}$&8977.0   &    0.0000179389\\
&$\beta_2KN_{III}$&8977.0   &    0.0000259348\\[2mm]
Zn
&$\alpha_2KL_{II}$&8615.78  &    0.14062\\
&$\alpha_1KL_{III}$&8638.86  &    0.273809\\
&$\beta_3KM_{II}$&9572.0   &    0.017389\\
&$\beta_KM_{III}$&9572.0   &    0.0339969\\
&$\beta_2KN_{II}$&9650.1   &    0.0000235079\\
&$\beta_2KN_{III}$&9650.1   &    0.0000339889\\[2mm]
\hline
\end{tabular}

\end{document}